\documentclass[aps,english,superscriptaddress,pra,twocolumn]{revtex4-1}
\usepackage{enumerate}
\usepackage{graphicx}
\usepackage{csquotes}
\usepackage{bbm}
\usepackage{natbib}
\usepackage[misc]{ifsym}
\usepackage{physics}
\usepackage{subfig}
\usepackage{multirow}
\usepackage{xcolor}
\allowdisplaybreaks[3]
\usepackage{ragged2e}
\DeclareCaptionJustification{justified}{\justifying}
\usepackage[font=small]{caption}

\usepackage{amssymb, amsbsy, amsthm}
\usepackage{times} 
\usepackage{empheq}

\begin{document} 
\title{Variational learning of integrated quantum photonic circuits}

\author{Hui Zhang}
\affiliation{Institute of Quantum Technologies (IQT), The Hong Kong Polytechnic University, Hong Kong}
\affiliation{Quantum Science and Engineering Centre (QSec), Nanyang Technological University, Singapore}
\author{Chengran Yang}
\affiliation{Centre for Quantum Technologies, National University of Singapore, 3 Science Drive 2, Singapore 117543}
\author{Wai-Keong Mok}
\affiliation{California Institute of Technology, Pasadena, CA 91125, USA}
\author{Lingxiao Wan}
\affiliation{Quantum Science and Engineering Centre (QSec), Nanyang Technological University, Singapore}
\author{Hong Cai}
\affiliation{Institute of Microelectronics, A*STAR (Agency for Science, Technology and Research), Singapore}
\author{Qiang Li}
\affiliation{Advanced Micro Foundry, 11 Science Park Rd, Singapore}
\author{Feng Gao}
\affiliation{Advanced Micro Foundry, 11 Science Park Rd, Singapore}
\author{Xianshu Luo}
\affiliation{Advanced Micro Foundry, 11 Science Park Rd, Singapore}
\author{Guo-Qiang Lo}
\affiliation{Advanced Micro Foundry, 11 Science Park Rd, Singapore}
\author{Lip Ket Chin}
\affiliation{Department of Electrical Engineering, City University of Hong Kong, Kowloon Tong, Hong Kong SAR}
\author{Yuzhi Shi}
\affiliation{Institute of Precision Optical Engineering, School of Physics Science and Engineering, Tongji University, Shanghai 200092, China}
\author{Jayne Thompson}
\email{thompson.jayne2@gmail.com (J.T)}
\affiliation{Institute of High Performance Computing, Agency for Science, Technology and Research, Singapore 138632}
\author{Mile Gu}
\email{gumile@ntu.edu.sg (M.G)}
\affiliation{Complexity Institute, Nanyang Technological University, Singapore}
\affiliation{Nanyang Quantum Hub, School of Physical and Mathematical Sciences, Nanyang Technological University, Singapore}
\author{Ai Qun Liu}
\email{eaqliu@ntu.edu.sg (A.Q.L)}
\affiliation{Quantum Science and Engineering Centre (QSec), Nanyang Technological University, Singapore}
\affiliation{Institute of Quantum Technologies (IQT), The Hong Kong Polytechnic University, Hong Kong}

\begin{abstract}
Integrated photonic circuits play a crucial role in implementing quantum information processing in the noisy intermediate-scale quantum (NISQ) era. Variational learning is a promising avenue that leverages classical optimization techniques to enhance quantum advantages on NISQ devices. However, most variational algorithms are circuit-model-based and encounter challenges when implemented on integrated photonic circuits, because they involve explicit decomposition of large quantum circuits into sequences of basic entangled gates, leading to an exponential decay of success probability due to the non-deterministic nature of photonic entangling gates. Here, we present a variational learning approach for designing quantum photonic circuits, which directly incorporates post-selection and elementary photonic elements into the training process. The complicated circuit is treated as a single nonlinear logical operator, and a unified design is discovered for it through variational learning. Engineering an integrated photonic chip with automated control, we adjust and optimize the internal parameters of the chip in real time for task-specific cost functions. We utilize a simple case of designing photonic circuits for a single ancilla CNOT gate with improved success rate to illustrate how our proposed approach works, and then apply the approach in the first demonstration of quantum stochastic simulation using integrated photonics.

\end{abstract}
\maketitle

\textbf{\textit{Introduction}} - Integrated quantum photonic circuits present a promising technological alternative for quantum information processing~\cite{silverstone2014chip, preble2015chip,paesani2020near,silverstone2015qubit,li2017chip,llewellyn2020chip,osada2019strongly, katsumi2019quantum}. They possess notable advantages, such as room temperature operation and miniaturization, whereby size and power consumption are reduced by orders of magnitude compared to superconducting systems~\cite{elshaari2020hybrid,zhou2023prospects,lu2021advances}. Notable achievements have recently been made within such platforms, including the realization of selected error-correcting codes~\cite{vigliar2021error,zhang2023encoding}, graph-related computations~\cite{arrazola2021quantum,bao2023very}, variational quantum eigensolvers~\cite{peruzzo2014variational} and quantum neural networks for hamiltonian learning~\cite{wang2017experimental}.

Photonic circuits exhibit a number of unique properties not shared by other circuit-based models of quantum computing. Examples include the relative ease of demonstration of quantum supremacy, thanks to the technological scalability of linear optics~\cite{spring2013boson,paesani2019generation,hoch2022reconfigurable}, and the capacity to prepare quantum operations in superposition due to the existence of the vacuum modes~\cite{zhou2011adding}. Meanwhile, photonic circuits also face unique challenges, as direct methods to implement entangling gates involve post-selection and are highly non-deterministic. As such, a naive translation of near-term circuit-based algorithms suffers exponentially diminishing success rates as we cascade such entangling operations~\cite{ralph2001simple,o2003demonstration,bharti2022noisy}. This is particularly pertinent in the NISQ (Noisy Intermediate-Scale Quantum) era that forbids fault-tolerant means of photonic computation~\cite{knill2001scheme} - severely limiting, for example, the circuit depth of experimentally realizable photonic circuits in variational settings~\cite{peruzzo2014variational, wang2017experimental}.

Here, we present an alternative means to design NISQ photonic circuits tailored to these unique quirks of photonic quantum computing. Instead of using circuit ansatzes that explicitly break a large $n$-qubit quantum circuit into sequences of elementary entangling gates, we consider a parametrization - with parameters quadratic in $n$ - that directly incorporates post-selection and the elementary optical elements in integrated photonics. Despite this efficient representation, results from Boson sampling~\cite{spring2013boson,paesani2019generation} guarantee we can access certain quantum operations that cannot be efficiently simulated classically.

\begin{figure*}[t]
\centering
\includegraphics[width=0.92\textwidth]{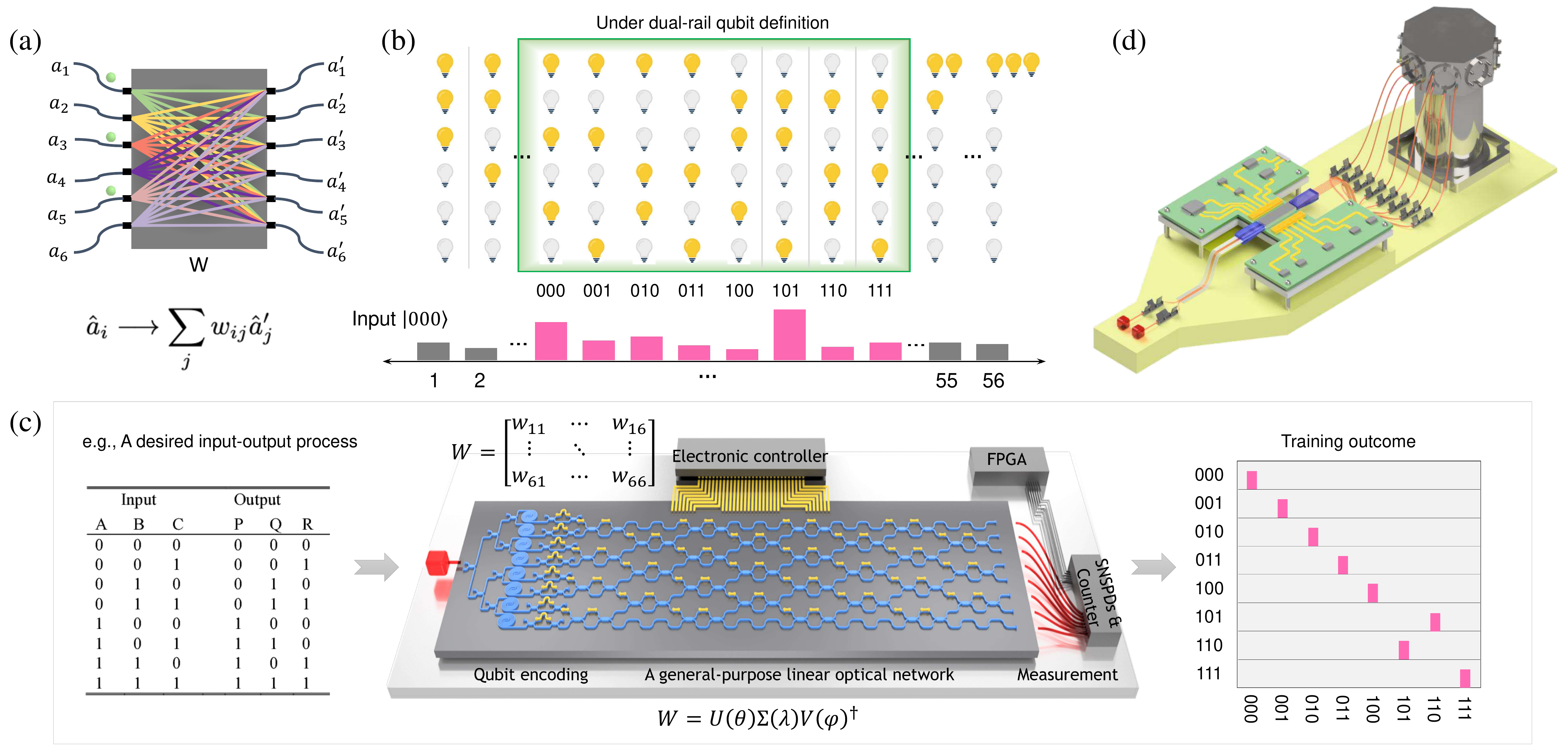}
\vspace{-0.3cm}
\caption{The framework of the variational learning of integrated quantum photonic circuits. (a) The quantum photonic circuit - an arbitrary complex-valued weight matrix $W$. Identical photons are incident and encoded in dual rail. (b) The output probability distribution of all the three-photon count events computed by $W$. Among these events, output states in the computational subspace (in the green box) are considered valid. (c) The mapping of the circuit design problem to variational learning involves training over $W$. Given a desired gate or input-output process, this training can drive the output probability distributions to the desired one, thereby achieving the training objective. (d) The control system of the on-chip training process.}
\label{fig:1}
\end{figure*}

We illustrate our technique via an integrated photonic circuit consisting of eight optical modes, that encode logical qubits through dual-rail encoding. We then engineered an automated control system that allows the adjustment of fundamental optical elements within the integrated circuit in real time. Coupling the control system with the performance of candidate integrated circuit designs then allowed us to automate the learning and design of integrated photonic circuits for specific tasks. We apply this system to two tasks: (i) designing a CNOT gate whose success probability is improved by an ancillary photon, illustrating the working principle of the proposed approach; and (ii) learning optical circuits to implement quantum interactions for quantum-enhanced stochastic simulation. Here, we implement variational learning directly on the chip, thus pioneering techniques to realize variational algorithms native to integrated photonics.

\textbf{\textit{Framework and Archetecture}} - 
We implement an $n$-qubit quantum gate through a linear optical network on a programmable photonic chip, as shown in \textbf{Fig.~\ref{fig:1}a}. 
We use the dual-rail encoding of photons, where a logical qubit is represented by a pair of adjacent waveguide path modes $(\hat{a}_{2i-1},\hat{a}_{2i})$, $i = 1, \ldots, n$. 
The logical $\ket{0}_i$ for $i$-th qubit means the $i$-th photon occurs at path mode $2i-1$, i.e., $\ket{0}_i \equiv \hat{a}_{2i-1}^\dag \ket{\text{vac}}$ while logical $\ket{1}_i \equiv \hat{a}_{2i}^\dag \ket{\text{vac}}$. $\ket{\text{vac}}$ denotes the vacuum state.
Certain logical unitary operator $\bar{U}$ - including some that cannot be efficiently simulated classically - is then realized via a generic complex matrix $W = (w_{ij})$ (non-unitary in general) acting on the path  mode operators $a_j$,
\begin{equation}
    \hat{a}_i  \rightarrow \sum_j w_{ij}(\theta) \hat{a}_j ^\prime 
\end{equation}
where $\hat{a}_i$ and $\hat{a}_j^\prime$ represents the input and output mode respectively. $W$ is programmable by the trainable parameters $\theta$. The elements of the unitary matrix $\bar{U}$ are permanent of the submatrix of matrix $W$~\cite{aaronson2011computational,broome2013photonic} (the computation of a 3-photon case is detailed in \textbf{Appendix A}).
To construct $\bar{U}$, we post-select those valid outputs where only a single photon appears on the adjacent waveguide mode $(\hat{a}_{2i-1},\hat{a}_{2i})$ (see \textbf{Fig.~\ref{fig:1}b}).

A $W$ matrix can be realized by a programmable quantum photonic chip consisting of beam splitters $\hat{U}_{\text{MMI}}$ and phase shifters $\hat{U}_{\text{PS}}$ ~\cite{reck1994experimental,shen2017deep} (see \textbf{Appendix B, C} for details) when its spectral norm $\Vert W\Vert\leq 1$. According to singular value decomposition, an arbitrary complex-valued matrix $W$ can be decomposed into two unitary matrices and a diagonal matrix as
\begin{equation}
    W = R_1\Sigma R_2^\dagger,
\end{equation}
where $R_1$ and $R_2^\dag$ are unitary matrices, and $\Sigma$ is a diagonal matrix $\Sigma = diag(\lambda_1,\cdots,\lambda_{2n})$ with singular values $\lambda_1 \geq \ldots \geq \lambda_{2n}$. $R_1$ and $R_2^\dag$ can be directly implemented by properly arranging the order of beam splitters and phase shifters and choosing the proper parameters $\theta$ for phase shifters.
When $\lambda_1 \leq 1$, the matrix $\Sigma$ can be realized by applying photon loss on each mode. Thus, we normalize $W$ matrix by dividing it by its spectral norm $\Vert W \Vert = \lambda_1$, denoted by $\tilde{W} = W/\Vert W \Vert$. The success probability of the post-selection-based transformation $W$ is $1/\|W\|^{2n}$~\cite{li2022quantum}.

\begin{figure*}[t]
\centering
\includegraphics[width=0.88\textwidth]{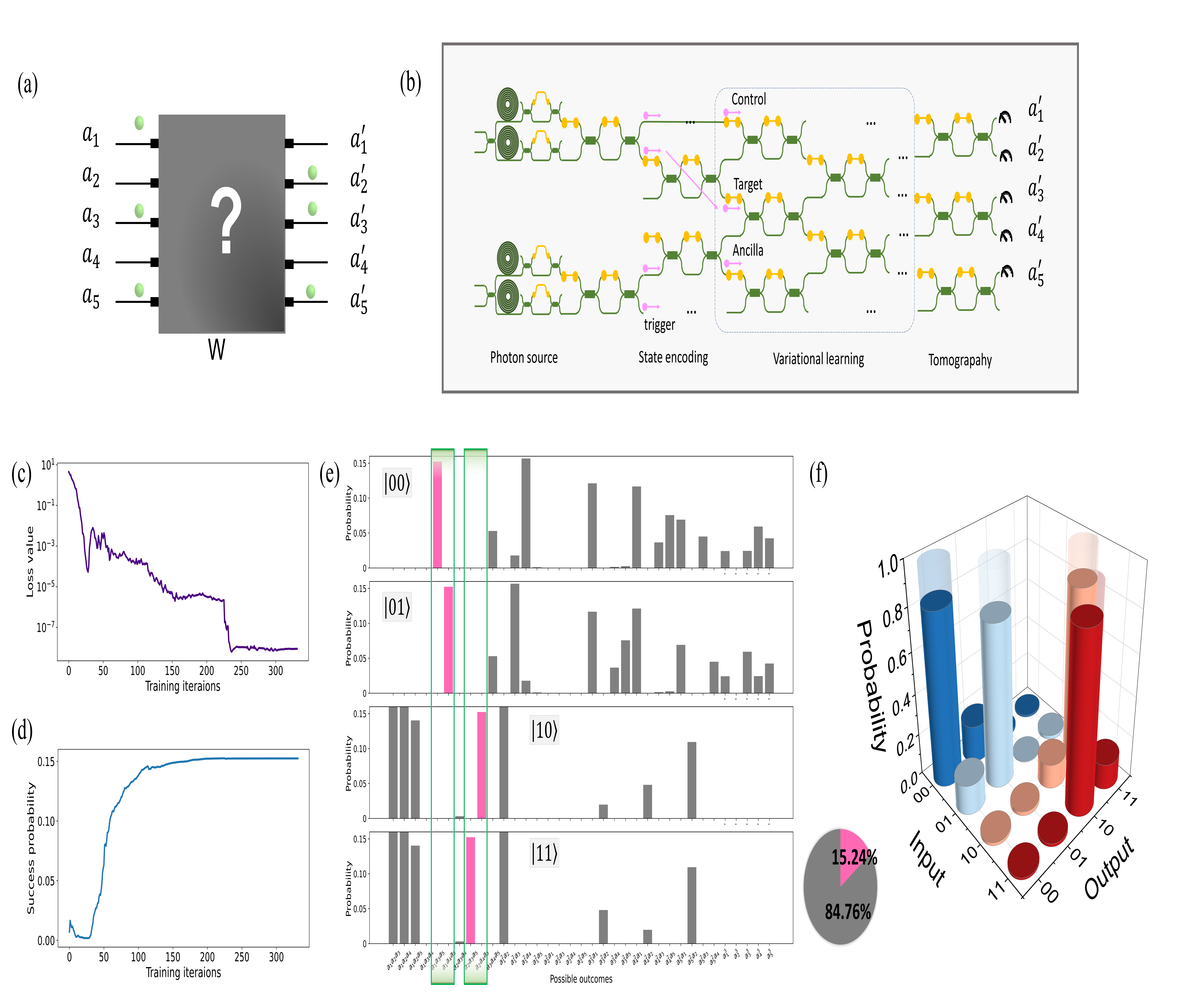}
\caption{The results of the single ancilla CNOT gate. (a,b) The $l_2$ loss and the success probability during training. (c) The probability distribution of all possible outcomes for the 3-photon case. (d) The experimental logical truth table of computational basis inputs to the CNOT gate.}
\label{fig:h}
\vspace{-0.3cm}
\end{figure*}

In the experiment, we realize a chip-integrated photonic circuit that includes the generation of degenerated photon pair sources and the linear optical network, as shown in \textbf{Fig.~\ref{fig:1}c}. The chip is fabricated on silicon-on-insulator platforms and is programmable using electronically controllable thermal-optic phase shifters. A dual-wavelength (1546.8 nm and 1553.2 nm) pump is coupled into the photonic chip, generating 1550 nm photons through the degenerated spontaneous four-wavelength mixing process. The photons are in the N00N state $\ket{\psi}=1/\sqrt{2}(\ket{02}+\ket{20})$ when only one pair photon term is considered. 
The visibility of the Hong-Ou-Mandel (HOM) interference between two pairs of photon sources is $0.852\pm0.065$ at a coincidence count of 2000 Hz for the two photon pairs. The deviations of HOM visibility from unit are primarily caused by reduced quantum interference between different pair creation events~\cite{tanida2012highly}, and higher-order terms in the photon generation process. Therefore, a trade-off exists between the visibility and the count rate by adjusting the pump power. After passing through the optical circuit, the photons are filtered by wavelength division multiplexing devices and detected by the superconducting nanowire single-photon detectors. 
The input photons are encoded in a dual-rail manner, occupying a total of 8 waveguide paths. Our chip can implement arbitrary complex-valued transformations $W$ on up to $4$ path mode operators $a_i$, thus facilitating the implementation of a generic logic unitary operator $\bar{U}$ on two qubits. 
Alternatively, we can implement unitary transformation $W$ on up to $8$ path mode operators $a_i$, thus restricting the implementation of any generic logical unitary operators. More details are provided in \textbf{Appendix D}.

As shown in \textbf{Fig.~\ref{fig:1}d}, an automated control system is engineered to adjust the control parameters $\theta$. This allows us to vary which $w_{ij}(\theta)$ is realized by our circuit in real-time, and thus easily implement a diverse array of potential unitaries. In particular, for each desired task, we then introduce an associated positive cost function
\begin{equation}
    \min_{\theta} f(\theta).
\end{equation}
The cost function value is zero if and only if the task is performed to perfection. Minimization of this cost function by varying $\theta$ then provides a means to discover optimal control parameters. This optimisation can be done using a variety of methods, including both offline on classical computer, or online by direct measurement-feedback on the integrated circuit. We illustrate the former in designing a single ancilla CNOT gate and the latter in quantum stochastic simulation.


\textbf{\textit{A single ancilla CNOT gate}} - As a simple illustration, we first showcase the discovery and implementation of a single ancilla CNOT gate with improved success probability. This case is intended to illustrate how our approach works, without focusing on benchmarking against previous demonstrations. The CNOT gate is a key component of quantum computation - and a particularly challenging element for photonics. With a CNOT gate, the logical state of the target qubit is flipped if the control qubit is in state 1 and left unchanged if the control qubit is in state 0. The CNOT logic are inherently nonlinear, so to be performed using simple linear optical elements, post-selection is required by designing specific configurations of the linear circuit to provide nonlinearity. Using linear optics, the success probability of implementing a CNOT gate cannot exceed $1/9$ without ancilla photons~\cite{shadbolt2012generating,ralph2002linear}. Such a CNOT gate has been demonstrated with a success probability of 1/9~\cite{carolan2015universal} in integrated photonic circuits. Meanwhile, in the presence of ancilla photons, a heralded CNOT gate can be achieved with a success probability of 1/16 using two separate ancilla photons~\cite{carolan2015universal}, and a single ancilla CNOT gate has been demonstrated with a success probability of 1/8~\cite{pittman2003experimental}.

Here we show the discovery of a single ancilla CNOT gate with an improved success probability. Our model consists of $5$ waveguides, the last of which is only used for the ancilla photon (see ~\textbf{Fig.~\ref{fig:h}a, b}). Two photons are used for dual rail encoding and are fed into the first four waveguides, whereas the third photon is fed into the last waveguide and only events with photon clicks at the ancilla port are accepted. Our target is to find a configuration of the programmable circuit that fulfills the logic of a CNOT gate, through variational learning. 
During training, from each assumed $W$ matrix on the programmable circuit, the underlying unitary operator $\bar{U}$ can be induced (see \textbf{Appendix E}). We numerically tune the elements of $W$ to minimize the $l_2$ norm between $\Bar{U}$ and the target CNOT $U_{\text{CNOT}}$, while maximizing the success probability. Thus the cost function is
\begin{equation}
    C = \|\bar{U} - U_{\text{CNOT}}\|_2 + \alpha (1- \frac{1}{\|W\|^6})
\end{equation}
where we set $\alpha =10^{-3}$ for tuning the parameter $W$.
As the number of iterations increases, the $l_2$ loss decreases to $10^{-7}$ while the success probability reaches $0.1524$, as shown in \textbf{Fig.~\ref{fig:h}a, b}. The $l_2$ loss of $10^{-7}$ means the underlying unitary $\bar{U}$ of $W$-matrix is almost the same as desired CNOT unitary.
We further implement the trained $W$ (see \textbf{Appendix E} for more details) in our photonic chip using $5$ of $8$ path modes. The measured logical truth table in the computational basis for the CNOT gate is depicted in \textbf{Fig.~\ref{fig:h}d}, with the ideal theoretical truth table overlaid. The mean statistical fidelity averaged over all computational inputs is 0.829$\pm$0.013. Although the trained $W$ theoretically allows for a CNOT logic with unit fidelity, in this multiphoton experiment, deviations from unit fidelity are primarily caused by imperfections in the photon source and the visibility of MZIs, and the thermal crosstalk, which prevents the implemented matrix from being ideal. Overall, our discovery of the photonic implementation through the variational learning approach improves the success probability to 0.1524. While modest, our methodology illustrates that our automated methods can design circuits that rival those designed by hand.


\textbf{\textit{Simulating stochastic process}} - Our next example is in quantum stochastic simulations, where quantum models can accelerate stochastic analysis~\cite{blank2021quantum}, generate futures in quantum superpositions~\cite{ghafari2019interfering,binder2018practical}, and do so while tracking fewer data than classically possible~\cite{gu2012quantum,mahoney2016occam}. A stochastic process generates output $x_t$, taken value from an alphabet, at each time step $t$.
Formally, a stochastic process is a probability distribution over some sequence of random variables $\{X_t\}$, representing the output of a stochastic system at each time step $t$. Taking $t = 0$ as the present, each instance of a stochastic process has a particular past $\overleftarrow{x} = \ldots x_{-2}x_{-1}$. 
A predictive model serves to encode this past $\overleftarrow{x}$ into some 
memory $M$, such that systematic actions on $M$ can generate the desired conditional future $\overrightarrow{x} = x_0x_1\ldots$ governed by $P(\overrightarrow{X} = \overrightarrow{x}|\overleftarrow{x})$. 


All classical predictive models are finite-state machines, with internal states $\{S_i\}$, and dynamics described by $P(x,S_j|S_i)$ -- the probability that a machine in state $S_i$ outputs $x$ and transitions to $S_j$~\cite{Crutchfield1989}. The memory cost of such a machine is then given by the Shannon entropy the distribution $\{p_i\}$, where $p_i$ is the probability that the machine is in state $S_i$ -- representing the amount of past information that such a machine needs to generate a statistically correct prediction about the future. The minimal memory needed is known as the statistical complexity, a fundamental measure of a process's internal structure that finds utility in diverse contexts~\cite{crutchfield2012between,shalizi2004quantifying,munoz2020general}.

\begin{figure}[t]
\centering
\includegraphics[width=0.40\textwidth]{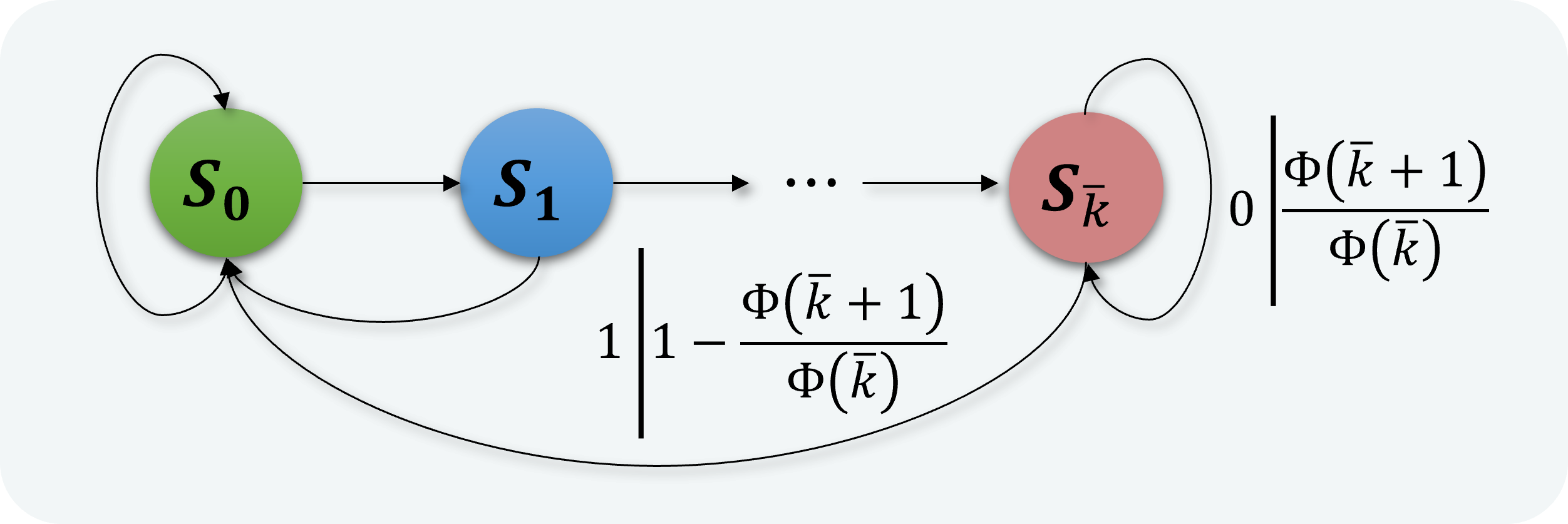}
\caption{Transition diagram of a renewal process. Renewal processes represent stochastic systems that output sequences of $0$'s (no tick) punctuated by $1$s (ticks). If the chance of ticking is independent of the number of previous $0$s, then a renewal process becomes Poisson and its simulation is memoryless. However, in more general scenarios, their simulation requires tracking the number of $0$s since the last tick and can thus scale without bounds. The dual Poisson process is a special class of renewal processes, where the probability of transition after $k$ $0$s is given by $1-\frac{\Phi(k+1)}{\Phi(k)}$, where $\Phi(k) = pq_1^k + (1-p)q_2^k$, and $p$, $q_1$ and $q_2$ are free parameters that lie between $0$ and $1$.}
\vspace{-0.3cm}
\label{fig:scheme}
\end{figure}

Quantum models encode past information into quantum states $\ket{\sigma_i}$ in place of classical states $S_i$~\cite{gu2012quantum, Aghamohammadi2017PRX, yang2018matrix}. At each time step, the quantum memory state $\ket{\sigma_i}$ is coupled with an ancilla register in the $\ket{0}$ state via a coupling unitary operator $U$. After the $U$-operation, measuring the ancilla register generates the output of the stochastic process and collapses the memory state for further simulation. Since all the past information is encoded in the quantum state $\ket{\sigma_i}$, the quantum memory is quantified by the von-Neumann entropy $C_q = H(\sum_i p_i \ket{\sigma_i}\bra{\sigma_i})$. In many situations, this memory can be lower than any classical counterpart~\cite{elliott2019extreme}. Meanwhile, the outputs of such quantum models generate all possible conditional futures in superposition. Previously, these advantages have been demonstrated in bulk optics~\cite{palsson2017experimentally,ghafari2019interfering,wu2023implementing}, but never within integrated photonics. Moreover, the breakdown of $U$ into optical elements was hand-designed.

\begin{figure}[t]
\centering
\includegraphics[width=0.45\textwidth]{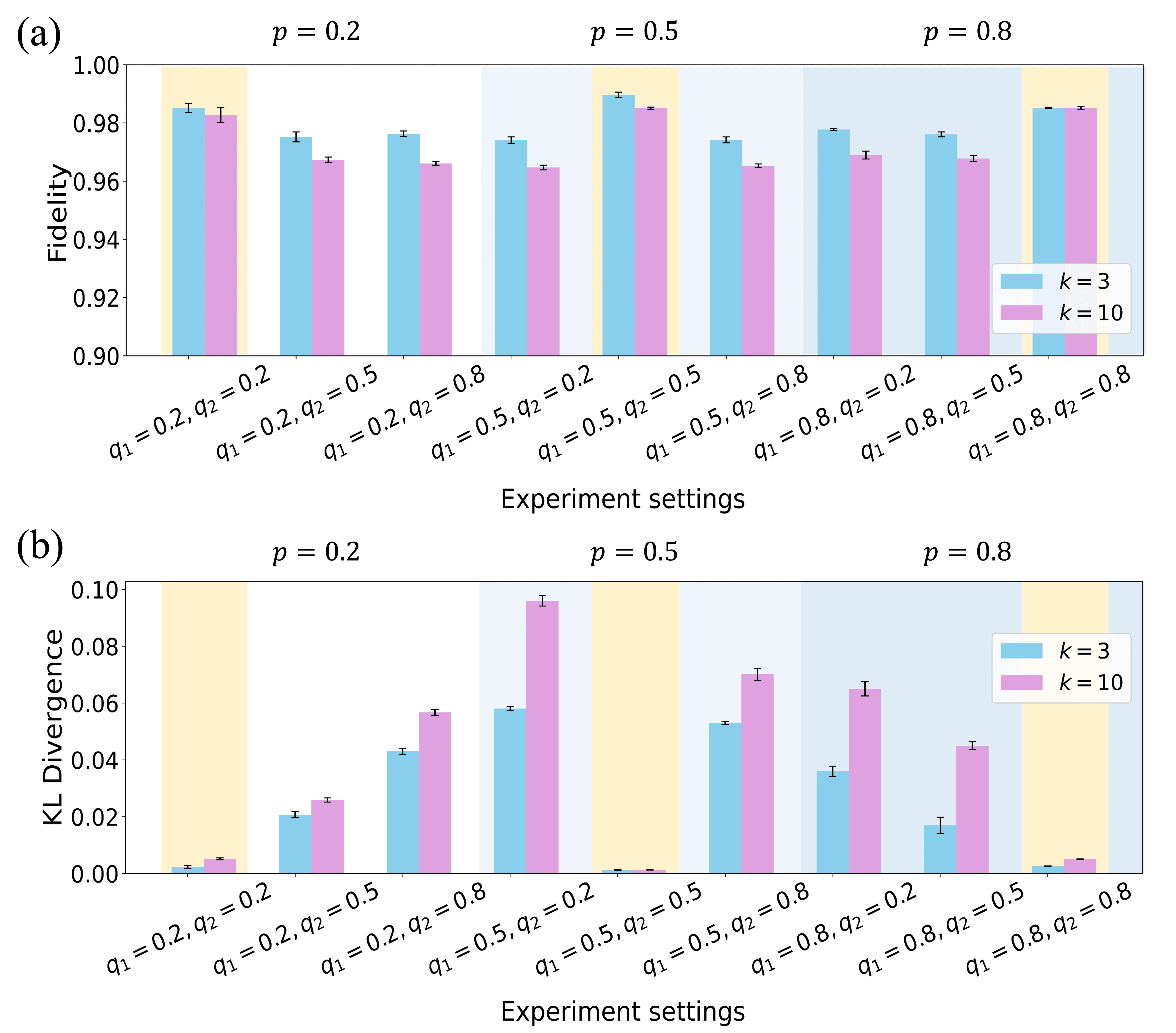}
\caption{Experimental results of dual Poisson process.
(a) The fidelity of memory state, for discrete $p$, $q_1$, and $q_2$ values. The average fidelity is 97.9\% for training (the simulation step $k=3$, blue) and 97.2\% for validation fidelities ($k=10$, pink).
(b) The KL divergence of the transition probability distribution. The model has the best performance for $q_1=q_2$, in the yellow area.}
\vspace{-0.3cm}
\label{fig:6}
\end{figure}

\begin{figure}[t]
\centering
\includegraphics[width=0.48\textwidth]{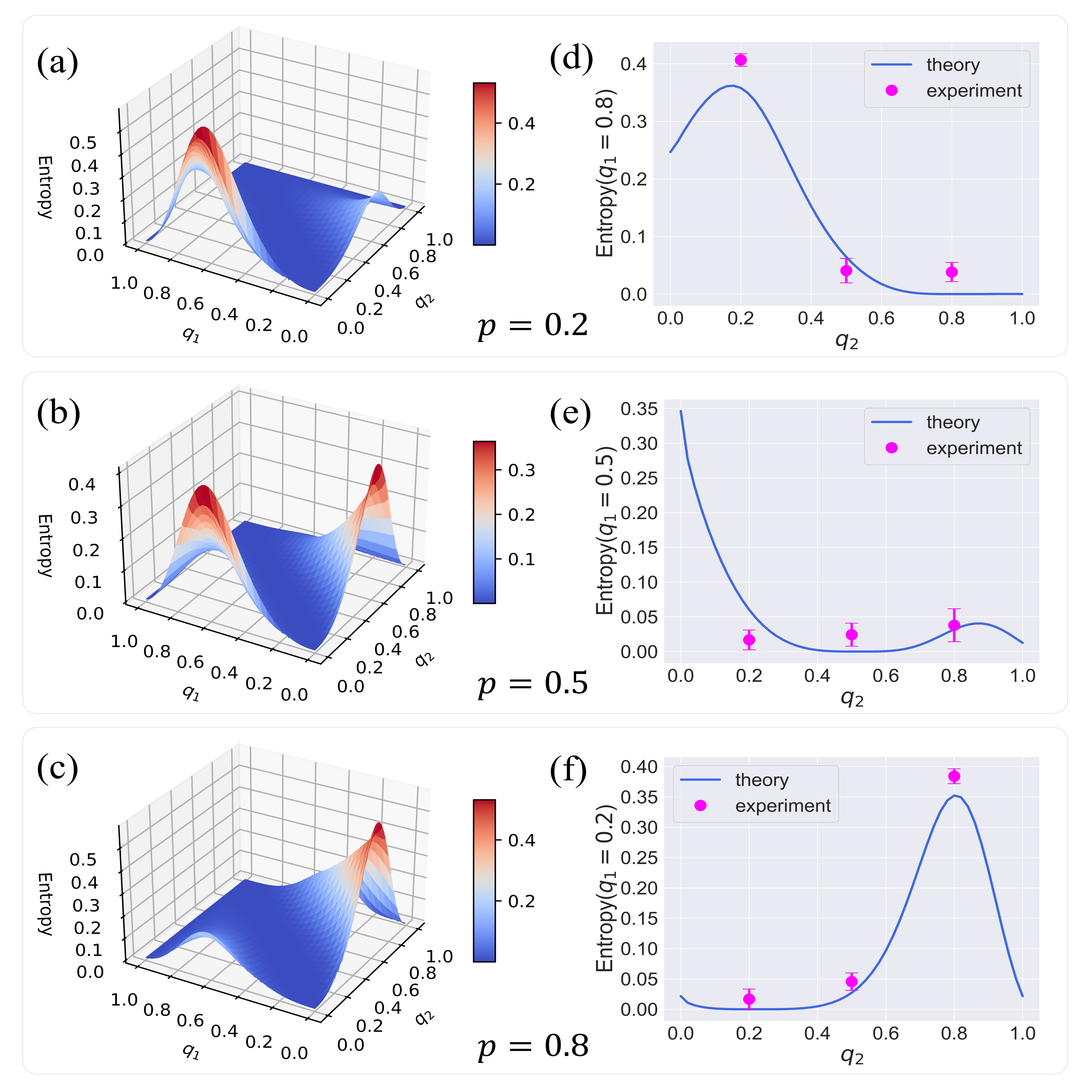}
\caption{(a)-(c) The theoretically calculated statistical complexity $C_q$ of the quantum simulator. Specific cross-sections with discrete parameter settings $(p, q_1, q_2)$, are truncated and experimentally probed, as shown in panels (d)-(e). The quantum statistical complexity (magenta dots) is computed according to the entropy of the reconstructed stationary states. The blue curves represent the theoretical $C_q$ truncated from (a)-(c). Uncertainties are estimated from the Poissonian distribution of photon counts.
}
\label{fig:qcentro}
\end{figure}

Our cost function consists of two parts: (i) the fidelity $f_1$ of the tomographic reconstructed memory state (conditioned on the ancilla outcome), 
\begin{equation}
    f_1 = \frac{1}{NL}\sum_{x,i} \left|\braket{\bar{ \sigma}_o}{\sigma_o}_{x,i}\right |^2
\end{equation} 
where $\ket{\sigma_o}_{x, i}$ denotes the state of the memory system after detecting an output $x$ given the input $\ket{\sigma_i}$, $N$ is the number of simulation steps, and $L$ is the alphabet size. $\ket{\bar{\sigma}_0}$ is the target output state.
(ii) the distance $f_2$ between the theoretical and experimental transition probability,
\begin{equation}
    f_2 = \frac{1}{N}\sum_{x,i} \sqrt{\left| P(x,S_j|S_i)-\bar{P} (x,S_j|S_i) \right|^2},
\end{equation}
where $\bar{P}$ is outcome probability obtained from experiment, $S_j$ is the casual states. 
We obtain the complete cost function by $f = (1-f_1) + \alpha f_2$, where $\alpha$ is a hyperparameter weighting $f_1$ and $f_2$.
We also calculate the Kullback Leibler (KL) divergence between theory/experimental transition probability distributions as a metric for the modeling accuracy (see definition in Appendix F). 

In this two-photon experiment, the training process involves executing the assumed chip configuration on the actual chip during each training iteration and collecting its responses through quantum state tomography. Cost function values are calculated according to the measurement results. Strategies are then generated according to the cost function value, using a training algorithm executed on a classical computer to adjust the programmable parameters on the chip. The training algorithm is gradient-free, following the genetic algorithm as previously reported~\cite{zhang2021efficient}. In short, we experiment with multiple sets of parameters on the chip and assess their cost function values. Parameters demonstrating good performance are chosen and utilized to generate the next generation of populations (i.e., sets of programmable parameters) for the subsequent training iteration, through genetic operators such as crossover, mutation, etc. In our implementation, each training iteration involves 30 populations that are uploaded to the chip. Considering that quantum state tomography is performed on the memory qubit conditioned on the ancilla qubit, repeated 30 times in each generation, and taking into account the algorithm processing time, the communication time between chip and processing unit, each iteration of the training process takes approximately 30 minutes. The entire variational learning task is planned to undergo 100 training iterations.

\begin{figure}[t]
\centering
\includegraphics[width=0.35\textwidth]{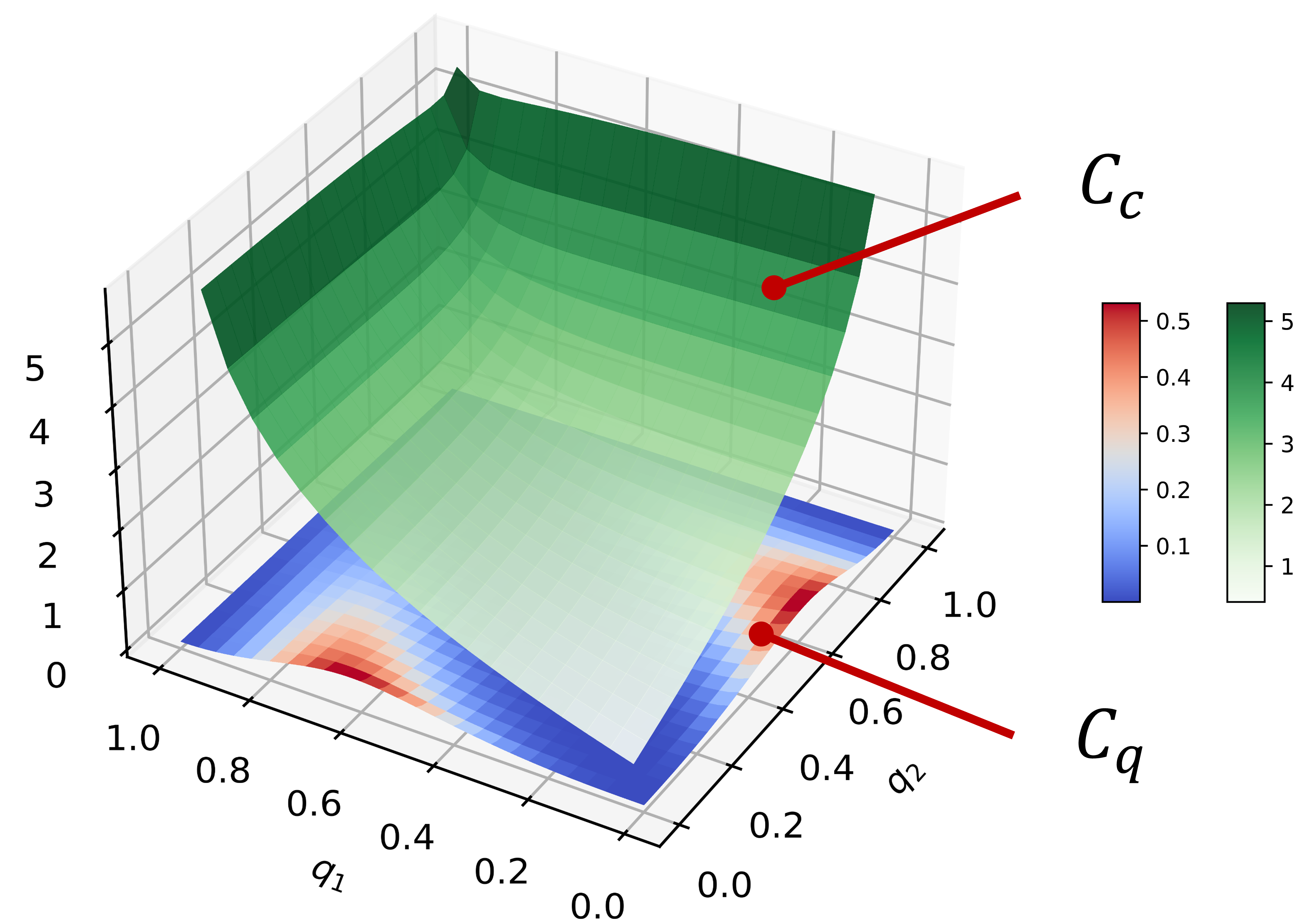}
\caption{The comparison of quantum entropy $C_q$ and classical entropy $C_c$.}
\label{fig:comp}
\end{figure}

Here, we train our integrated photonic circuits to simulate the dual Poisson process, a sub-class of renewal processes \textbf{(Fig.~\ref{fig:scheme}}) for which quantum memory advantage is known to scale without bound~\cite{elliott2019extreme}. Our goal is to obtain the $U$-operator to realize the dual Poisson process (see details of $U$ in Appendix G). The experiment is conducted for discrete $p$ values of 0.2, 0.5, and 0.8, with three sets of $(q_1, q_2)$ settings under each $p$. \textbf{Figures~\ref{fig:6}a, b} show the fidelity of the memory state and the KL divergence of transition probability by measuring the ancilla state. The theoretical quantum entropy and classical entropy calculated are shown in \textbf{Figs.~\ref{fig:qcentro}a-c} and \textbf{Figs.~A2a-c} (Appendix H), respectively. Cross-sections are truncated and probed experimentally, as shown in \textbf{Figs.~\ref{fig:qcentro}d-f}. The quantum entropy (magenta dots) is obtained experimentally according to the reconstructed stationary states. The results proves the reduced entropy $C_q<C_c$ for any parameters $p$, $q_1$ and $q_2$. The comparison of $C_q$ and $C_c$ for the case when $p=0.5$ is visualized in Fig.~\ref{fig:comp}.

\textbf{\textit{Conclusion}} - 
We present and experimentally demonstrate a variational approach to quantum photonic circuit design, offering a systematic and efficient methodology for optimizing circuit configurations. By engineering a photonic chip coupled to an automated control system, we are able to adjust and optimize the internal parameters of the chip in real time for task-specific cost functions. We demonstrate the versatility of this approach on two distinct tasks, the engineering of a single ancilla CNOT Gate on photonic circuits, and the first demonstration of quantum-enhanced stochastic simulation using integrated photonics.

Our variational approach has a significant advantage over the direct implementation of variational circuit algorithms intended for use in other computing platforms. Notably, the standard approach involves explicitly decomposing large quantum circuits into sequences of basic entangled gates -- which would cause exponential die-off in success probabilities due to the non-deterministic nature of photonic entangling gates. In contrast, our variational techniques directly incorporate post-selection and the fundamental photonics elements into the training process. Notably, it does not mean we can overcome the probabilistic nature of implementing logic gates on photonic circuits. Instead, we are proposing a method for designing the post-selection configurations of the chip. Therefore, for complex problems composed of a series of probabilistic gates in a standard circuit model, we treat it as a single nonlinear logical operator and generate a unified design for it, which conforms to the characteristics that probabilistic photonic quantum operators cannot be effectively cascaded. Our advances thus provide a promising pathway for the development of variational algorithms native to the unique advantages and challenges of integrated photonic circuits.

\textbf{Acknowledgement} --
This work is supported by the Singapore Ministry of Education Tier 3 grant (MOE2017-T3-1-001), the Agency for Science, Technology and Research (A*STAR) under
its QEP2.0 programme (NRF2021-QEP2-02-P06) and the Singapore Ministry of Education Tier 1 Grant RG77/22. 

\textbf{Author Contribution} -- H.Z. and C.R.Y. conceived the ideas. H.Z. designed the chip and built the experimental setup with assistance from L.X.W. Q.L., F.G., H.C., X.S.L., and G.Q.L. fabricated the silicon photonic chip. H.Z. performed the experiments, analyzed the data, and wrote the manuscript with contributions from all co-authors. C.R.Y. contributed to writing the theoretical part of the paper. All authors participated in discussing the experimental results. J.T. and M.G. revised the paper. J.T., M.G., and A.Q.L. supervised and coordinated all the work. L.K.C., Y.Z.S., and A.Q.L. provided valuable comments on the manuscript.

\bibliographystyle{apsrev4-1}
\bibliography{n}


\end{document}


\title{\large Appendix for \\
Variational learning of integrated quantum photonic circuits}

\author{Hui Zhang}
\author{Chengran Yang}
\author{Wai-Keong Mok}
\author{Lingxiao Wan}
\author{Hong Cai}
\author{Qiang Li}
\author{Feng Gao}
\author{Xianshu Luo}
\author{Guo-Qiang Lo}
\author{Lip Ket Chin}
\author{Yuzhi Shi}
\author{Jayne Thompson}
\email{thompson.jayne2@gmail.com (J.T)}
\author{Mile Gu}
\email{gumile@ntu.edu.sg (M.G)}
\author{Ai Qun Liu}
\email{eaqliu@ntu.edu.sg (A.Q.L)}
\maketitle

\section*{Appendix A: Complimentary information of framework}
We take a three-qubit case as an example. Three photons are encoded on six path waveguide modes. The computational input states are $\ket{000}$,$\ket{001}$,$\ket{010}$, $\ket{011}$,$\ket{100}$,$\ket{101}$,$\ket{110}$, and $\ket{111}$. They can be written as
\begin{equation}
\begin{aligned}
    &\ket{000}: \hat{a}_{1in}^\dag\hat{a}_{3in}^\dag\hat{a}_{5in}^\dag \ket{\text{vac}}, 
    \ket{001}: \hat{a}_{1in}^\dag\hat{a}_{3in}^\dag\hat{a}_{6in}^\dag \ket{\text{vac}},\\
    &\ket{010}: \hat{a}_{1in}^\dag\hat{a}_{4in}^\dag\hat{a}_{5in}^\dag \ket{\text{vac}}, 
    \ket{011}: \hat{a}_{1in}^\dag\hat{a}_{4in}^\dag\hat{a}_{6in}^\dag \ket{\text{vac}}\\
    &\ket{100}: \hat{a}_{2in}^\dag\hat{a}_{3in}^\dag\hat{a}_{5in}^\dag \ket{\text{vac}}, 
    \ket{101}: \hat{a}_{2in}^\dag\hat{a}_{3in}^\dag\hat{a}_{6in}^\dag \ket{\text{vac}},\\
    &\ket{110}: \hat{a}_{2in}^\dag\hat{a}_{4in}^\dag\hat{a}_{5in}^\dag \ket{\text{vac}}, 
    \ket{111}: \hat{a}_{2in}^\dag\hat{a}_{4in}^\dag\hat{a}_{6in}^\dag \ket{\text{vac}}\\
\end{aligned}
\end{equation}.
The $W$-matrix performs the evolution between the input and output modes as:
\begin{align}
    \begin{bmatrix}
    \hat{a}_{1in}^\dag\\
    \hat{a}_{2in}^\dag\\
    \hat{a}_{3in}^\dag\\
    \hat{a}_{4in}^\dag\\
    \hat{a}_{5in}^\dag\\
    \hat{a}_{6in}^\dag\\
    \end{bmatrix} \rightarrow
    \begin{bmatrix}
    w_{11} &w_{12} &w_{13} &w_{14} &w_{15} &w_{16}\\
    w_{21} &w_{22} &w_{23} &w_{24} &w_{25} &w_{26}\\
    w_{31} &w_{32} &w_{33} &w_{34} &w_{35} &w_{36}\\
    w_{41} &w_{42} &w_{43} &w_{44} &w_{45} &w_{46}\\
    w_{51} &w_{52} &w_{53} &w_{54} &w_{55} &w_{56}\\
    w_{61} &w_{62} &w_{63} &w_{64} &w_{65} &w_{66}\\
    \end{bmatrix}
    \begin{bmatrix}
    \hat{a}_{1out}^\dag\\
    \hat{a}_{2out}^\dag\\
    \hat{a}_{3out}^\dag\\
    \hat{a}_{4out}^\dag\\
    \hat{a}_{5out}^\dag\\
    \hat{a}_{6out}^\dag\\ 
    \end{bmatrix}
\end{align}
So the evolution of the computation basis $\ket{000}$ are:
\begin{align}
\small
\begin{aligned}
\ket{000}_{in} &= \hat{a}_{1in}^\dag\hat{a}_{3in}^\dag\hat{a}_{5in}^\dag \ket{\text{vac}}\\
&\rightarrow(w_{11}\hat{a}_{1out}^\dag+w_{12}\hat{a}_{2out}^\dag+w_{13}\hat{a}_{3out}^\dag+w_{14}\hat{a}_{4out}^\dag+w_{15}\hat{a}_{5out}^\dag+w_{16}\hat{a}_{6out}^\dag)\\&(w_{31}\hat{a}_{1out}^\dag+w_{32}\hat{a}_{2out}^\dag+w_{33}\hat{a}_{3out}^\dag+w_{34}\hat{a}_{4out}^\dag+w_{35}\hat{a}_{5out}^\dag+w_{36}\hat{a}_{6out}^\dag)\\&(w_{51}\hat{a}_{1out}^\dag+w_{52}\hat{a}_{2out}^\dag+w_{53}\hat{a}_{3out}^\dag+w_{54}\hat{a}_{4out}^\dag+w_{55}\hat{a}_{5out}^\dag+w_{56}\hat{a}_{6out}^\dag)\ket{vac}\\
&\xrightarrow{post-select} (w_{11}w_{33}w_{55}+w_{13}w_{35}w_{51}+w_{15}w_{31}w_{53}+w_{15}w_{33}w_{51}+w_{13}w_{31}w_{55}+w_{11}w_{35}w_{53})\hat{a}_{1out}^\dag\hat{a}_{3out}^\dag\hat{a}_{5out}^\dag\ket{vac}\\
&+(w_{11}w_{33}w_{56}+w_{13}w_{36}w_{51}+w_{16}w_{31}w_{53}+w_{16}w_{33}w_{51}+w_{13}w_{31}w_{56}+w_{11}w_{36}w_{53})\hat{a}_{1out}^\dag\hat{a}_{3out}^\dag\hat{a}_{6out}^\dag\ket{vac}\\
&+(w_{11}w_{34}w_{55}+w_{14}w_{35}w_{51}+w_{15}w_{31}w_{54}+w_{15}w_{34}w_{51}+w_{14}w_{31}w_{55}+w_{11}w_{35}w_{54})\hat{a}_{1out}^\dag\hat{a}_{4out}^\dag\hat{a}_{5out}^\dag\ket{vac}\\
&+(w_{11}w_{34}w_{56}+w_{14}w_{36}w_{51}+w_{16}w_{31}w_{54}+w_{16}w_{34}w_{51}+w_{14}w_{31}w_{56}+w_{11}w_{36}w_{54})\hat{a}_{1out}^\dag\hat{a}_{4out}^\dag\hat{a}_{6out}^\dag\ket{vac}\\
&+(w_{12}w_{33}w_{55}+w_{13}w_{35}w_{52}+w_{15}w_{32}w_{53}+w_{15}w_{33}w_{52}+w_{13}w_{32}w_{55}+w_{12}w_{35}w_{53})\hat{a}_{2out}^\dag\hat{a}_{3out}^\dag\hat{a}_{5out}^\dag\ket{vac}\\
&+(w_{12}w_{33}w_{56}+w_{13}w_{36}w_{52}+w_{16}w_{32}w_{53}+w_{16}w_{33}w_{52}+w_{13}w_{32}w_{56}+w_{12}w_{36}w_{53})\hat{a}_{2out}^\dag\hat{a}_{3out}^\dag\hat{a}_{6out}^\dag\ket{vac}\\
&+(w_{12}w_{34}w_{55}+w_{14}w_{35}w_{52}+w_{15}w_{32}w_{54}+w_{15}w_{34}w_{52}+w_{14}w_{32}w_{55}+w_{12}w_{35}w_{54})\hat{a}_{2out}^\dag\hat{a}_{4out}^\dag\hat{a}_{5out}^\dag\ket{vac}\\
&+(w_{12}w_{34}w_{56}+w_{14}w_{36}w_{52}+w_{16}w_{32}w_{54}+w_{16}w_{34}w_{52}+w_{14}w_{32}w_{56}+w_{12}w_{36}w_{54})\hat{a}_{2out}^\dag\hat{a}_{4out}^\dag\hat{a}_{6out}^\dag\ket{vac}\\
&=\rm perm(\begin{bmatrix}
        w_{11} &w_{13} &w_{15}\\
        w_{31} &w_{33} &w_{35}\\
        w_{51} &w_{53} &w_{55}\\
    \end{bmatrix})\ket{000}_{out}
    +\rm perm(\begin{bmatrix}
        w_{11} &w_{13} &w_{16}\\
        w_{31} &w_{33} &w_{36}\\
        w_{51} &w_{53} &w_{56}\\
    \end{bmatrix})\ket{001}_{out}
    +\rm perm(\begin{bmatrix}
        w_{11} &w_{14} &w_{15}\\
        w_{31} &w_{34} &w_{35}\\
        w_{51} &w_{54} &w_{55}\\
    \end{bmatrix})\ket{010}_{out}\\
    &+\rm perm(\begin{bmatrix}
        w_{11} &w_{14} &w_{16}\\
        w_{31} &w_{34} &w_{36}\\
        w_{51} &w_{54} &w_{56}\\
    \end{bmatrix})\ket{011}_{out}
    +\rm perm(\begin{bmatrix}
        w_{12} &w_{13} &w_{15}\\
        w_{32} &w_{33} &w_{35}\\
        w_{52} &w_{53} &w_{55}\\
    \end{bmatrix})\ket{100}_{out}
    +\rm perm(\begin{bmatrix}
        w_{12} &w_{13} &w_{16}\\
        w_{32} &w_{33} &w_{36}\\
        w_{52} &w_{53} &w_{56}\\
    \end{bmatrix})\ket{101}_{out}\\
    &+\rm perm(\begin{bmatrix}
        w_{12} &w_{14} &w_{15}\\
        w_{32} &w_{34} &w_{35}\\
        w_{52} &w_{54} &w_{55}\\
    \end{bmatrix})\ket{110}_{out}
    +\rm perm(\begin{bmatrix}
        w_{12} &w_{14} &w_{16}\\
        w_{32} &w_{34} &w_{36}\\
        w_{52} &w_{54} &w_{56}\\
    \end{bmatrix})\ket{111}_{out}
\end{aligned}
\end{align}
and  example for input basis of $\ket{001}$,
\begin{align}
\small
\begin{aligned}
\ket{001}_{in} &= \hat{a}_{1in}^\dag\hat{a}_{3in}^\dag\hat{a}_{6in}^\dag \ket{\text{vac}}\\
&\rightarrow(w_{11}\hat{a}_{1out}^\dag+w_{12}\hat{a}_{2out}^\dag+w_{13}\hat{a}_{3out}^\dag+w_{14}\hat{a}_{4out}^\dag+w_{15}\hat{a}_{5out}^\dag+w_{16}\hat{a}_{6out}^\dag)\\&(w_{31}\hat{a}_{1out}^\dag+w_{32}\hat{a}_{2out}^\dag+w_{33}\hat{a}_{3out}^\dag+w_{34}\hat{a}_{4out}^\dag+w_{35}\hat{a}_{5out}^\dag+w_{36}\hat{a}_{6out}^\dag)\\&(w_{61}\hat{a}_{1out}^\dag+w_{62}\hat{a}_{2out}^\dag+w_{63}\hat{a}_{3out}^\dag+w_{64}\hat{a}_{4out}^\dag+w_{65}\hat{a}_{5out}^\dag+w_{66}\hat{a}_{6out}^\dag)\ket{vac}\\
&\xrightarrow{post-select} (w_{11}w_{33}w_{65}+w_{13}w_{35}w_{61}+w_{15}w_{31}w_{63}+w_{15}w_{33}w_{61}+w_{13}w_{31}w_{65}+w_{11}w_{35}w_{63})\hat{a}_{1out}^\dag\hat{a}_{3out}^\dag\hat{a}_{5out}^\dag\ket{vac}\\
&+(w_{11}w_{33}w_{66}+w_{13}w_{36}w_{61}+w_{16}w_{31}w_{63}+w_{16}w_{33}w_{61}+w_{13}w_{31}w_{66}+w_{11}w_{36}w_{63})\hat{a}_{1out}^\dag\hat{a}_{3out}^\dag\hat{a}_{6out}^\dag\ket{vac}\\
&+(w_{11}w_{34}w_{65}+w_{14}w_{35}w_{61}+w_{15}w_{31}w_{64}+w_{15}w_{34}w_{61}+w_{14}w_{31}w_{65}+w_{11}w_{35}w_{64})\hat{a}_{1out}^\dag\hat{a}_{4out}^\dag\hat{a}_{5out}^\dag\ket{vac}\\
&+(w_{11}w_{34}w_{66}+w_{14}w_{36}w_{61}+w_{16}w_{31}w_{64}+w_{16}w_{34}w_{61}+w_{14}w_{31}w_{66}+w_{11}w_{36}w_{64})\hat{a}_{1out}^\dag\hat{a}_{4out}^\dag\hat{a}_{6out}^\dag\ket{vac}\\
&+(w_{12}w_{33}w_{65}+w_{13}w_{35}w_{62}+w_{15}w_{32}w_{63}+w_{15}w_{33}w_{62}+w_{13}w_{32}w_{65}+w_{12}w_{35}w_{63})\hat{a}_{2out}^\dag\hat{a}_{3out}^\dag\hat{a}_{5out}^\dag\ket{vac}\\
&+(w_{12}w_{33}w_{66}+w_{13}w_{36}w_{62}+w_{16}w_{32}w_{63}+w_{16}w_{33}w_{62}+w_{13}w_{32}w_{66}+w_{12}w_{36}w_{63})\hat{a}_{2out}^\dag\hat{a}_{3out}^\dag\hat{a}_{6out}^\dag\ket{vac}\\
&+(w_{12}w_{34}w_{65}+w_{14}w_{35}w_{62}+w_{15}w_{32}w_{64}+w_{15}w_{34}w_{62}+w_{14}w_{32}w_{65}+w_{12}w_{35}w_{64})\hat{a}_{2out}^\dag\hat{a}_{4out}^\dag\hat{a}_{5out}^\dag\ket{vac}\\
&+(w_{12}w_{34}w_{66}+w_{14}w_{36}w_{62}+w_{16}w_{32}w_{64}+w_{16}w_{34}w_{62}+w_{14}w_{32}w_{66}+w_{12}w_{36}w_{64})\hat{a}_{2out}^\dag\hat{a}_{4out}^\dag\hat{a}_{6out}^\dag\ket{vac}\\
&=\rm perm(\begin{bmatrix}
        w_{11} &w_{13} &w_{15}\\
        w_{31} &w_{33} &w_{35}\\
        w_{61} &w_{63} &w_{65}\\
    \end{bmatrix})\ket{000}_{out}
    +\rm perm(\begin{bmatrix}
        w_{11} &w_{13} &w_{16}\\
        w_{31} &w_{33} &w_{36}\\
        w_{61} &w_{63} &w_{66}\\
    \end{bmatrix})\ket{001}_{out}
    +\rm perm(\begin{bmatrix}
        w_{11} &w_{14} &w_{15}\\
        w_{31} &w_{34} &w_{35}\\
        w_{61} &w_{64} &w_{65}\\
    \end{bmatrix})\ket{010}_{out}\\
    &+\rm perm(\begin{bmatrix}
        w_{11} &w_{14} &w_{16}\\
        w_{31} &w_{34} &w_{36}\\
        w_{61} &w_{64} &w_{66}\\
    \end{bmatrix})\ket{011}_{out}
    +\rm perm(\begin{bmatrix}
        w_{12} &w_{13} &w_{15}\\
        w_{32} &w_{33} &w_{35}\\
        w_{62} &w_{63} &w_{65}\\
    \end{bmatrix})\ket{100}_{out}
    +\rm perm(\begin{bmatrix}
        w_{12} &w_{13} &w_{16}\\
        w_{32} &w_{33} &w_{36}\\
        w_{62} &w_{63} &w_{66}\\
    \end{bmatrix})\ket{101}_{out}\\
    &+\rm perm(\begin{bmatrix}
        w_{12} &w_{14} &w_{15}\\
        w_{32} &w_{34} &w_{35}\\
        w_{62} &w_{64} &w_{65}\\
    \end{bmatrix})\ket{110}_{out}
    +\rm perm(\begin{bmatrix}
        w_{12} &w_{14} &w_{16}\\
        w_{32} &w_{34} &w_{36}\\
        w_{62} &w_{64} &w_{66}\\
    \end{bmatrix})\ket{111}_{out}
\end{aligned}
\end{align}
The evolution can be performed to other computational bases in this way. Thus, the 64 elements of the underlying $\bar{U}$ can be derived from the $W$-matrix as
\begin{equation}
    \bar{U}_{ij}=\rm perm\left(\begin{bmatrix}
        w_{s_i(1),s_j(1)} &w_{s_i(1),s_j(2)} &w_{s_i(1),s_j(3)}\\
        w_{s_i(2),s_j(1)} &w_{s_i(2),s_j(2)} &w_{s_i(2),s_j(3)}\\
        w_{s_i(3),s_j(1)} &w_{s_i(3),s_j(2)} &w_{s_i(3),s_j(3)}\\
    \end{bmatrix}\right),
\label{eq: 10}
\end{equation}
where $s_i$ and $s_j$ ($i,j=1,\cdots,8$) are the computational bases. $s_1=(1,3,5), s_2=(1,3,6),s_3=(1,4,5),s_4=(1,4,6),s_5=(2,3,5),s_6=(2,3,6),s_7=(2,4,5),s_8=(2,4,6)$.

\section*{Appendix B: Photonic operators}
Consider a photonic chip with $K$ paths, which constitutes $K$ optical modes. 
The state of the system is described by creation/annihilation operators $\hat{a}_i^\dagger /\hat{a}_i^\dagger$ on each of the $K$ modes
\begin{equation}
\begin{split}
        \ket{\psi} &= \sum_{n_1\cdots n_K} A_{n_1,\cdots, n_K}\ket{n_1\cdots n_K} \\
                   &= \sum_{n_1\cdots n_K} A_{n_1,\cdots, n_K}\frac{(\hat{a}_1^\dagger)^{n_1}}{\sqrt{n_1!}}\cdots \frac{(\hat{a}_K^\dagger)^{n_K}}{\sqrt{n_K!}}\ket{n_1\cdots n_K}
\end{split}
\end{equation}
where $n_i$ are photon numbers in $i^{\text{th}}$mode.
A type of lossless evolution (photon number preserving) is a passive linear transformation,
\begin{equation}\label{eq:linear transform}
    \hat{a}_i \rightarrow \sum_j w_{ij}\hat{a}_j^{\prime}
\end{equation}
where $w_{ij}$ is an arbitrary complex-valued matrix~\cite{vanloock2010Optical}, $\hat{a}_i$ and $\hat{a}_j^\prime$ represents the input and output mode respectively. We list several widely used passive linear transformations, such as beam splitters $\hat{U}_{\rm MMI}$ (or MMI), and phase shifters $\bar{U}_{\rm PS}$,
\begin{equation}
     \hat{U}_{\rm MMI} = \frac{1}{\sqrt{2}}\begin{bmatrix}
    1 &i \\
    i &1 
    \end{bmatrix}, \quad 
    \hat{U}_{\rm PS}(\theta)=\begin{bmatrix}
    e^{i\theta} &0\\
    0 &1
    \end{bmatrix}.
\end{equation}
Our chip consists of a number of unit blocks, which is implemented by a Mach-Zehnder interferometer consisting of two 50:50 directional couplers, preceded by a phase shift at one input port
\begin{equation*}
\begin{aligned}
    \hat{U}_{\rm MZI}&=\hat{U}_{\rm MMI}\hat{U}_{\rm PS}(\theta)\hat{U}_{\rm MMI}\hat{U}_{\rm PS}(\phi)\\
    &=ie^{i\frac{\theta}{2}}\begin{bmatrix}
    e^{i\phi}\sin(\frac{\theta}{2}) &e^{i\phi}\cos(\frac{\theta}{2}) \\
    \cos(\frac{\theta}{2}) &-\sin(\frac{\theta}{2})
    \end{bmatrix}.
\end{aligned}
\end{equation*}
In the SVD decomposition of the complex-valued matrix $W = U\Sigma V^\dagger$, the unitary part ($U$, $V$) of the chip corresponds to a passive linear transformation on creation/annihilation operators. The diagonal part ($\Sigma$) introduces photon loss on each of the modes, corresponding to applying a diagonal matrix to the creation/annihilation operators.

\section*{Appendix C: Dual-rail encoding and quantum model implementation in the chip}
Consider two optical modes with creation/annihilation operators $\hat{a}_1^\dagger,\hat{a}_2^\dagger / \hat{a}_1,\hat{a}_2$.
The basis of a logical qubit is defined as 
\begin{equation}
    \begin{split}
        \ket{0} &= \hat{a}_1^\dagger \ket{\text{vac}}\\
        \ket{1} &= \hat{a}_2^\dagger \ket{\text{vac}}
    \end{split}
\end{equation}
For example, in simulating the stochastic process, our chip consists of four input modes with creation/annihilation operators $\hat{a}_i^\dagger/\hat{a}_i$, every two modes of four is regarded as a qubit, as suggested by dual-rail encoding. The upper two modes ($\hat{a}_1/\hat{a}_2$) represent the memory qubit, while the lower two modes $\hat{a}_3, \hat{a}_4$ represent the output system. At the beginning of the process, two photons are fed into the chip. 
The memory photon is in the superposition of mode 1 and mode 2.
\begin{equation}
     \ket{\sigma_i} = c_0\ket{1000}_p + c_1\ket{0100}_p = (c_0\hat{a}_1^\dagger + c_1\hat{a}_2^\dagger) \ket{\text{vac}} 
\end{equation}
The output photon is created at mode $3$, which is is designated computational basis state $|0\rangle$,
\begin{equation}
   \ket{0} = \ket{0010}_p = \hat{a}_3^\dagger \ket{\text{vac}}.
\end{equation}
The chip produces the evolution of the mode operators,
\begin{equation}
    \hat{a}_i  \rightarrow \sum_j w_{ij}(\bar{\theta}) \hat{a}_j ^\prime 
\end{equation}
where $W=[w_{ij}]$ denotes a complex matrix, and $\bar{\theta}$ are a vector of parameters.
After the detection of photons, we post-select those results, that are compatible with dual-rail encoding. Namely, we accept the outcomes spanned on the following basis
\begin{equation}
\hat{a}_{i_1}^\dagger \hat{a}_{i_2}^\dagger \ket{\text{vac}},~~i_1\in\{1,2\}~~\text{and}~~i_2 \in \{3,4\}      
\end{equation}
We do the state tomography of the memory qubit. 

\section{Appendix D: Experimental setup}
\begin{figure*}[t]
\centering
\includegraphics[width=0.65\textwidth]{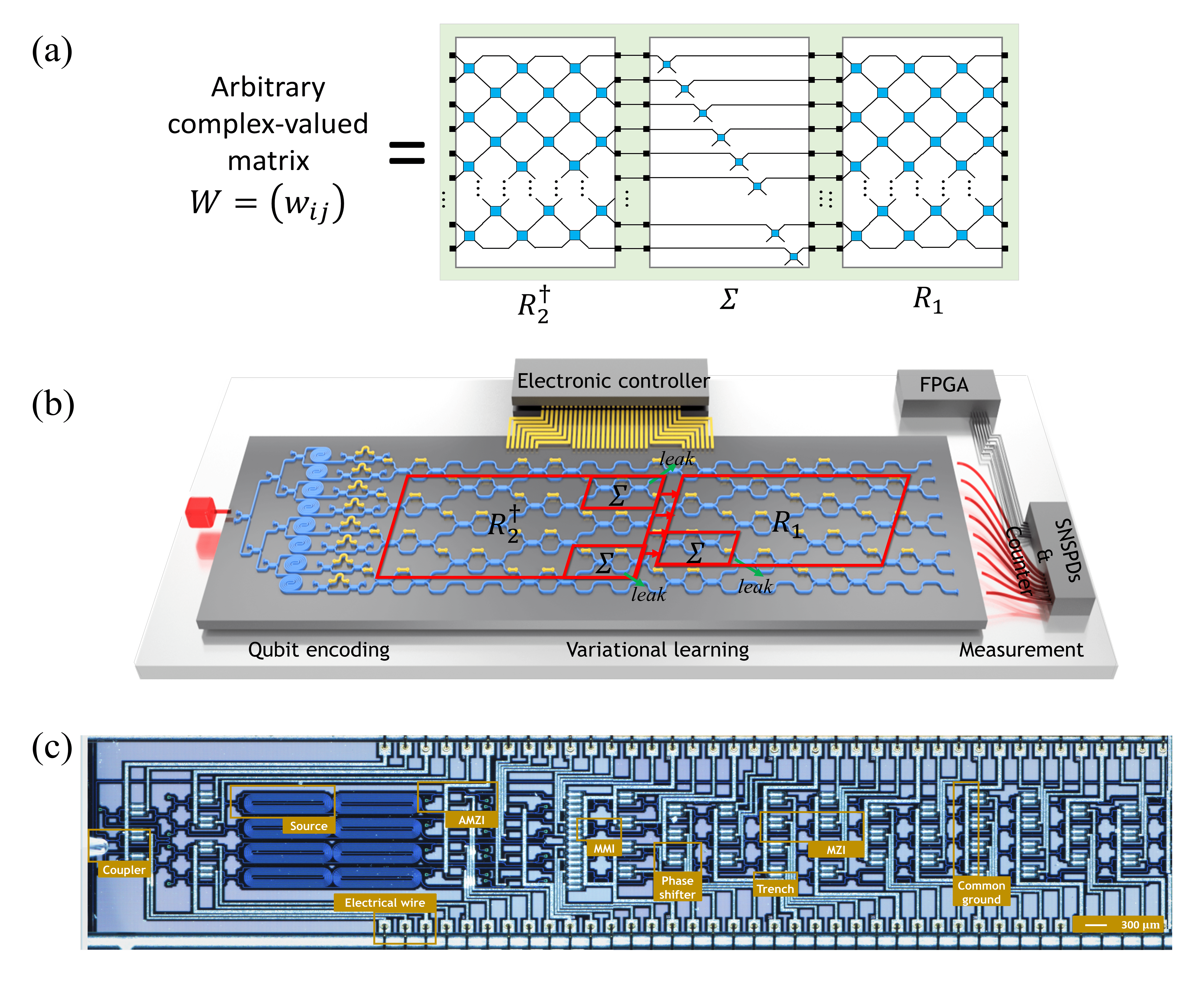}
\caption{(a) The singular value decomposition (SVD) of a complex-valued matrix. (b) The realization of a four-mode complex-valued transformation $W$ with the SVD structure, enabling a generic logic unitary operator $\bar{U}$ on two qubits. (c) The micrograph of the chip. }
\label{fig:SVD}
\vspace{-0.5cm}
\end{figure*}

Our framework trains over $W$, a complex-valued matrix implementable on the chip after performing singular value decomposition (SVD), as illustrated in Fig. \ref{fig:SVD}a. Our chip is an eight-mode linear optical circuit that can accommodate a complex-valued $W$ with up to four path modes, corresponding to the SVD decomposition results $W = R_1 \Sigma R_2^\dag$, as shown in Fig. \ref{fig:SVD}b. The micrograph of the fabricated chip is shown in Fig. \ref{fig:SVD}c, where photonic components like photon sources and adjustable phase shifter are marked.  The total loss of each photon from generation to detection sums up to 16 dB, which is consistent with a coincidences-to-singles ratio of 2.3\%. The main sources of optical losses include the coupling loss of 4.8 dB, the waveguide propagation loss of 4.6 dB (1.3 cm spirals and 1 cm straight waveguides at 2 dB/cm), the MZI loss of 3 dB (0.15 dB per MMI and there are around 20 MMIs on the optical path), and the detection loss of 4 dB.

\begin{figure*}[t]
\centering
\includegraphics[width=0.55\textwidth]{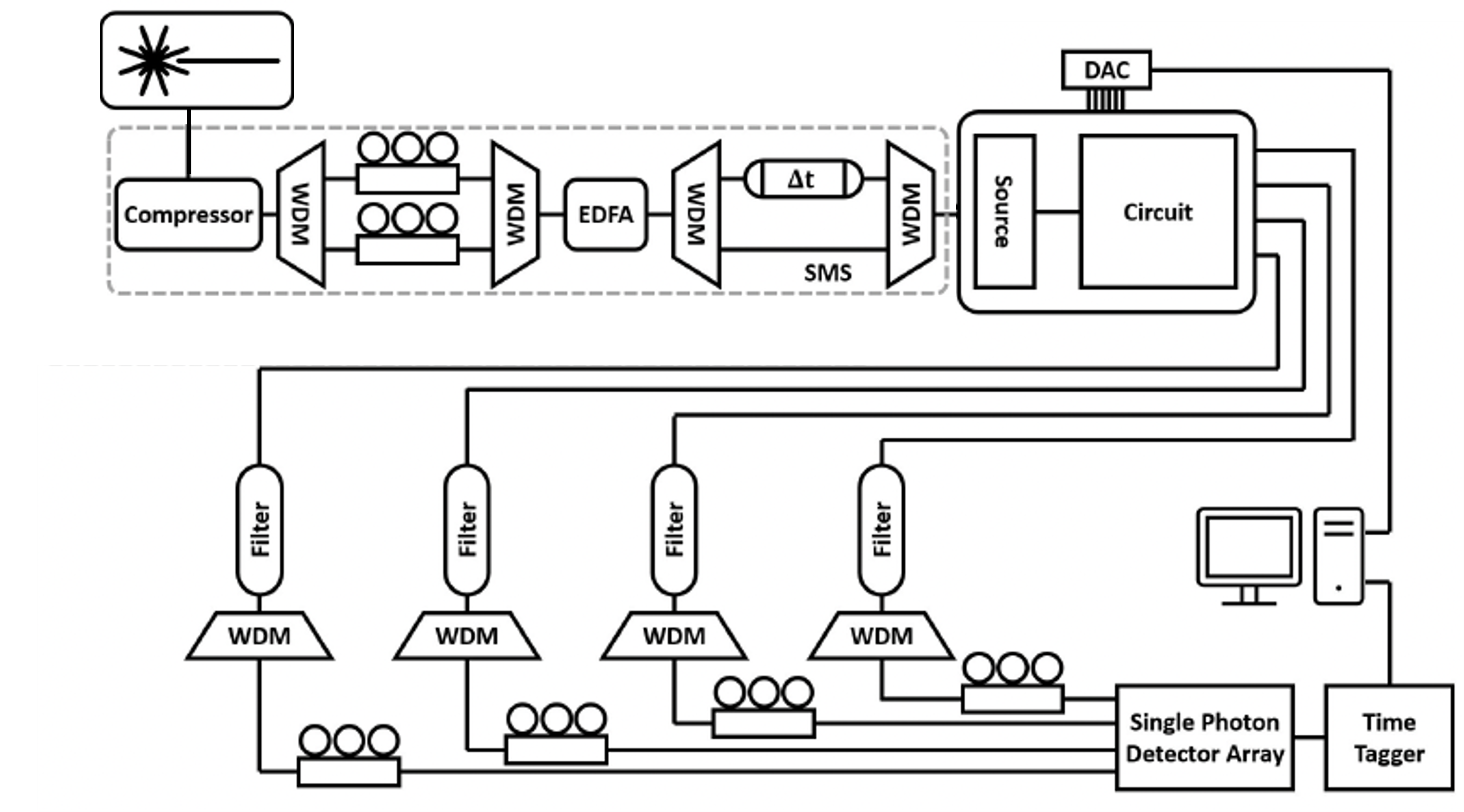}
\caption{The experimental setup for the integrated photonic circuit experiment, including a detailed photon generation process.}
\label{fig:setup}
\vspace{-0.5cm}
\end{figure*}

The experimental setup is shown in Fig. \ref{fig:setup}. The pump laser is generated from an ultrafast optical clock device with a repetition rate of 500 MHz, a central wavelength of 1550.116 nm, and a bandwidth of 1.9 nm. We generate a pair of identical photons on chip using a degenerate spontaneous four-wave mixing process. The pulsed laser first goes through a compressor to expand the bandwidth to about 10 nm. Then, two pump wavelengths are selected with a 100 G WDM (Wavelength Division Multiplexing) device and recombined into a single-mode fiber via another WDM. The laser power is then amplified with an Erbium-doped fiber amplifier (EDFA), since the wavelength selection process suppresses most parts of the spectra, thus keeping limited power. However, the EDFA would increase the background light again, so another pair of WDMs is used to remove the noise. A tunable delay line is used to balance the optical path difference of two pump wavelengths.

After coupling the dual pump wavelengths to the chip and characterizing the photon sources, the computational tasks are performed. The chip is controlled by analog signals transformed from digital signals by the digital-to-analog converter (DAC) and from the training algorithm. After computation, photons are coupled out of the chip and detected by single-photon detectors. The photon signals are then converted to electrical signals and processed by a time tagger and a classical unit. During the variational learning process, the processing of electrical signals, running the training algorithm, and outputting control digital signals are sequentially executed in each training iteration.

\begin{figure*}[t]
\centering
\includegraphics[width=0.5\textwidth]{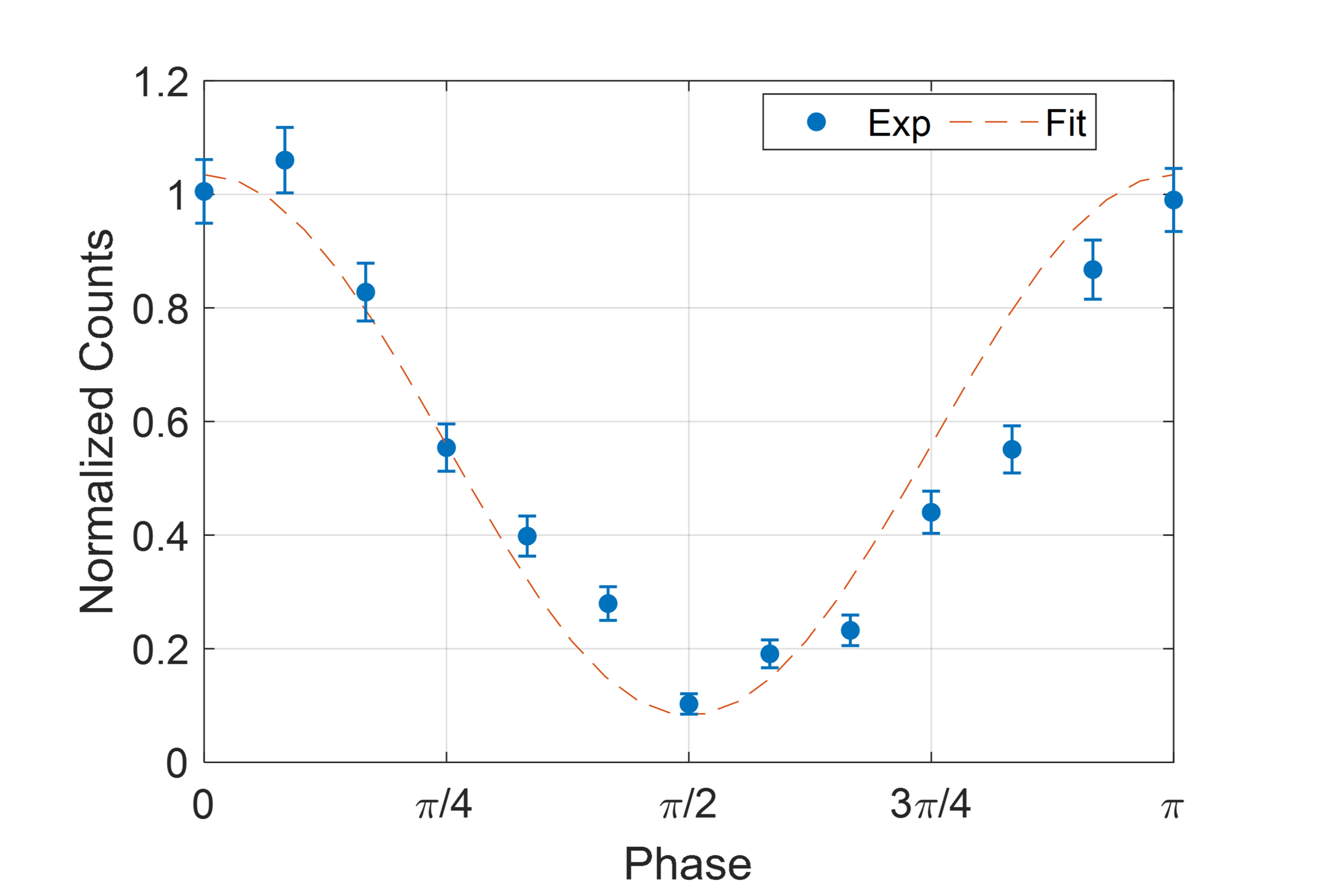}
\caption{ The HOM interference.}
\label{fig:source}
\vspace{-0.5cm}
\end{figure*}

In the CNOT case, two pairs of identical photons are needed. Three of them are used for the CNOT implementation, while the remaining single photon is used for heralding. We test the Hong-Ou-Mandel (HOM) interference between the two pairs of photon sources, obtaining a visibility of $0.852\pm0.065$, as shown in Fig. \ref{fig:source}b.

\section*{Appendix E: The ancilla-assited CNOT}
In this case, we have three photons and five path modes. Two photons are used as the control/target qubits of the CNOT gate, and the other photon is used as an ancilla. The computational input states are $\ket{00}$,$\ket{01}$,$\ket{10}$, and $\ket{11}$. They can be written as
\begin{equation}
\begin{aligned}
    &\ket{00}: \hat{a}_{1in}^\dag\hat{a}_{3in}^\dag\hat{a}_{5in}^\dag \ket{\text{vac}}, 
    \ket{01}: \hat{a}_{1in}^\dag\hat{a}_{4in}^\dag\hat{a}_{5in}^\dag \ket{\text{vac}},\\
    &\ket{10}: \hat{a}_{2in}^\dag\hat{a}_{3in}^\dag\hat{a}_{5in}^\dag \ket{\text{vac}}, 
    \ket{11}: \hat{a}_{2in}^\dag\hat{a}_{4in}^\dag\hat{a}_{5in}^\dag \ket{\text{vac}}
\end{aligned}
\end{equation}, respectively.
The $W$-matrix is a $5\times5$ matrix, performing the evolution between the input and output modes, which can be given by:
\begin{equation}
    \begin{bmatrix}
    \hat{a}_{1in}^\dag\\
    \hat{a}_{2in}^\dag\\
    \hat{a}_{3in}^\dag\\
    \hat{a}_{4in}^\dag\\
    \hat{a}_{5in}^\dag\\
    \end{bmatrix} \rightarrow
    \begin{bmatrix}
    w_{11} &w_{12} &w_{13} &w_{14} &w_{15}\\
    w_{21} &w_{22} &w_{23} &w_{24} &w_{25}\\
    w_{31} &w_{32} &w_{33} &w_{34} &w_{35}\\
    w_{41} &w_{42} &w_{43} &w_{44} &w_{45}\\
    w_{51} &w_{52} &w_{53} &w_{54} &w_{55}\\
    \end{bmatrix}
    \begin{bmatrix}
    \hat{a}_{1out}^\dag\\
    \hat{a}_{2out}^\dag\\
    \hat{a}_{3out}^\dag\\
    \hat{a}_{4out}^\dag\\
    \hat{a}_{5out}^\dag\\ 
    \end{bmatrix}
\end{equation}
So the evolution of the four computation bases are:
\begin{align}
\begin{aligned}
\ket{00}_{in} &= \hat{a}_{1in}^\dag\hat{a}_{3in}^\dag\hat{a}_{5in}^\dag \ket{\text{vac}}\\
&\rightarrow(w_{11}\hat{a}_{1out}^\dag+w_{12}\hat{a}_{2out}^\dag+w_{13}\hat{a}_{3out}^\dag+w_{14}\hat{a}_{4out}^\dag+w_{15}\hat{a}_{5out}^\dag)(w_{31}\hat{a}_{1out}^\dag+w_{32}\hat{a}_{2out}^\dag+w_{33}\hat{a}_{3out}^\dag+w_{34}\hat{a}_{4out}^\dag+w_{35}\hat{a}_{5out}^\dag)\\&(w_{51}\hat{a}_{1out}^\dag+w_{52}\hat{a}_{2out}^\dag+w_{53}\hat{a}_{3out}^\dag+w_{54}\hat{a}_{4out}^\dag+w_{55}\hat{a}_{5out}^\dag)\ket{vac}\\
&\xrightarrow{post-select} (w_{11}w_{33}w_{55}+w_{13}w_{35}w_{51}+w_{15}w_{31}w_{53}+w_{15}w_{33}w_{51}+w_{13}w_{31}w_{55}+w_{11}w_{35}w_{53})\hat{a}_{1out}^\dag\hat{a}_{3out}^\dag\hat{a}_{5out}^\dag\ket{vac}\\
&+(w_{11}w_{34}w_{55}+w_{14}w_{35}w_{51}+w_{15}w_{31}w_{54}+w_{15}w_{34}w_{51}+w_{14}w_{31}w_{55}+w_{11}w_{35}w_{54})\hat{a}_{1out}^\dag\hat{a}_{4out}^\dag\hat{a}_{5out}^\dag\ket{vac}\\
&+(w_{12}w_{33}w_{55}+w_{13}w_{35}w_{52}+w_{15}w_{32}w_{53}+w_{15}w_{33}w_{52}+w_{13}w_{32}w_{55}+w_{12}w_{35}w_{53})\hat{a}_{2out}^\dag\hat{a}_{3out}^\dag\hat{a}_{5out}^\dag\ket{vac}\\
&+(w_{12}w_{34}w_{55}+w_{14}w_{35}w_{52}+w_{15}w_{32}w_{54}+w_{15}w_{34}w_{52}+w_{14}w_{32}w_{55}+w_{12}w_{35}w_{54})\hat{a}_{2out}^\dag\hat{a}_{4out}^\dag\hat{a}_{5out}^\dag\ket{vac}\\
&=\rm perm(\begin{bmatrix}
        w_{1,1} &w_{1,3} &w_{1,5}\\
        w_{3,1} &w_{3,3} &w_{3,5}\\
        w_{5,1} &w_{5,3} &w_{5,5}\\
    \end{bmatrix})\ket{00}_{out}+\rm perm(\begin{bmatrix}
        w_{1,1} &w_{1,4} &w_{1,5}\\
        w_{3,1} &w_{3,4} &w_{3,5}\\
        w_{5,1} &w_{5,4} &w_{5,5}\\
    \end{bmatrix})\ket{01}_{out}+\\
    &\hspace{0.5 cm}\rm perm(\begin{bmatrix}
        w_{1,2} &w_{1,3} &w_{1,5}\\
        w_{3,2} &w_{3,3} &w_{3,5}\\
        w_{5,2} &w_{5,3} &w_{5,5}\\
    \end{bmatrix})\ket{10}_{out}+ \rm perm(\begin{bmatrix}
        w_{1,2} &w_{1,4} &w_{1,5}\\
        w_{3,2} &w_{3,4} &w_{3,5}\\
        w_{5,2} &w_{5,4} &w_{5,5}\\
    \end{bmatrix})\ket{11}_{out}
\end{aligned}
\end{align}
and similarly
\begin{equation}
\begin{aligned}
\ket{01}_{in} &= \hat{a}_{1in}^\dag\hat{a}_{4in}^\dag\hat{a}_{5in}^\dag \ket{\text{vac}}\\
&\rightarrow(w_{11}\hat{a}_{1out}^\dag+w_{12}\hat{a}_{2out}^\dag+w_{13}\hat{a}_{3out}^\dag+w_{14}\hat{a}_{4out}^\dag+w_{15}\hat{a}_{5out}^\dag)(w_{41}\hat{a}_{1out}^\dag+w_{42}\hat{a}_{2out}^\dag+w_{43}\hat{a}_{3out}^\dag+w_{44}\hat{a}_{4out}^\dag+w_{45}\hat{a}_{5out}^\dag)\\&(w_{51}\hat{a}_{1out}^\dag+w_{52}\hat{a}_{2out}^\dag+w_{53}\hat{a}_{3out}^\dag+w_{54}\hat{a}_{4out}^\dag+w_{55}\hat{a}_{5out}^\dag)\ket{vac}\\
&\xrightarrow{post-select} \rm perm(\begin{bmatrix}
        w_{11} &w_{13} &w_{15}\\
        w_{41} &w_{43} &w_{45}\\
        w_{51} &w_{53} &w_{55}\\
    \end{bmatrix})\ket{00}_{out}+\rm perm(\begin{bmatrix}
        w_{11} &w_{14} &w_{15}\\
        w_{41} &w_{44} &w_{45}\\
        w_{51} &w_{54} &w_{55}\\
    \end{bmatrix})\ket{01}_{out}+\\
    &\hspace{1.9 cm}\rm perm(\begin{bmatrix}
        w_{12} &w_{13} &w_{15}\\
        w_{42} &w_{43} &w_{45}\\
        w_{52} &w_{53} &w_{55}
    \end{bmatrix})\ket{10}_{out}+ \rm perm(\begin{bmatrix}
        w_{12} &w_{14} &w_{15}\\
        w_{42} &w_{44} &w_{45}\\
        w_{52} &w_{54} &w_{55}\\
    \end{bmatrix})\ket{11}_{out}
\end{aligned}
\end{equation}

\begin{equation}
\begin{aligned}
\ket{10}_{in} &= \hat{a}_{2in}^\dag\hat{a}_{3in}^\dag\hat{a}_{5in}^\dag \ket{\text{vac}}\\
&\rightarrow(w_{21}\hat{a}_{1out}^\dag+w_{22}\hat{a}_{2out}^\dag+w_{23}\hat{a}_{3out}^\dag+w_{24}\hat{a}_{4out}^\dag+w_{25}\hat{a}_{5out}^\dag)(w_{31}\hat{a}_{1out}^\dag+w_{32}\hat{a}_{2out}^\dag+w_{33}\hat{a}_{3out}^\dag+w_{34}\hat{a}_{4out}^\dag+w_{35}\hat{a}_{5out}^\dag)\\&(w_{51}\hat{a}_{1out}^\dag+w_{52}\hat{a}_{2out}^\dag+w_{53}\hat{a}_{3out}^\dag+w_{54}\hat{a}_{4out}^\dag+w_{55}\hat{a}_{5out}^\dag)\ket{vac}\\
&\xrightarrow{post-select} \rm perm(\begin{bmatrix}
        w_{21} &w_{23} &w_{25}\\
        w_{31} &w_{33} &w_{35}\\
        w_{51} &w_{53} &w_{55}\\
    \end{bmatrix})\ket{00}_{out}+\rm perm(\begin{bmatrix}
        w_{21} &w_{24} &w_{25}\\
        w_{31} &w_{34} &w_{35}\\
        w_{51} &w_{54} &w_{55}\\
    \end{bmatrix})\ket{01}_{out}+\\
    &\hspace{1.9 cm}\rm perm(\begin{bmatrix}
        w_{22} &w_{23} &w_{25}\\
        w_{32} &w_{33} &w_{35}\\
        w_{52} &w_{53} &w_{55}\\
    \end{bmatrix})\ket{10}_{out}+ \rm perm(\begin{bmatrix}
        w_{22} &w_{24} &w_{25}\\
        w_{32} &w_{34} &w_{35}\\
        w_{52} &w_{54} &w_{55}\\
    \end{bmatrix})\ket{11}_{out}
\end{aligned}
\end{equation}

\begin{equation}
\begin{aligned}
\ket{11}_{in} &= \hat{a}_{2in}^\dag\hat{a}_{4in}^\dag\hat{a}_{5in}^\dag \ket{\text{vac}}\\
&\rightarrow(w_{21}\hat{a}_{1out}^\dag+w_{22}\hat{a}_{2out}^\dag+w_{23}\hat{a}_{3out}^\dag+w_{24}\hat{a}_{4out}^\dag+w_{25}\hat{a}_{5out}^\dag)(w_{41}\hat{a}_{1out}^\dag+w_{42}\hat{a}_{2out}^\dag+w_{43}\hat{a}_{3out}^\dag+w_{44}\hat{a}_{4out}^\dag+w_{45}\hat{a}_{5out}^\dag)\\&(w_{51}\hat{a}_{1out}^\dag+w_{52}\hat{a}_{2out}^\dag+w_{53}\hat{a}_{3out}^\dag+w_{54}\hat{a}_{4out}^\dag+w_{55}\hat{a}_{5out}^\dag)\ket{vac}\\
&\xrightarrow{post-select} \rm perm(\begin{bmatrix}
        w_{21} &w_{23} &w_{25}\\
        w_{41} &w_{43} &w_{45}\\
        w_{51} &w_{53} &w_{55}\\
    \end{bmatrix})\ket{00}_{out}+\rm perm(\begin{bmatrix}
        w_{21} &w_{24} &w_{25}\\
        w_{41} &w_{44} &w_{45}\\
        w_{51} &w_{54} &w_{55}\\
    \end{bmatrix})\ket{01}_{out}+\\
    &\hspace{1.9 cm}\rm perm(\begin{bmatrix}
        w_{22} &w_{23} &w_{25}\\
        w_{42} &w_{43} &w_{45}\\
        w_{52} &w_{53} &w_{55}\\
    \end{bmatrix})\ket{10}_{out}+ \rm perm(\begin{bmatrix}
        w_{22} &w_{24} &w_{25}\\
        w_{42} &w_{44} &w_{45}\\
        w_{52} &w_{54} &w_{55}\\
    \end{bmatrix})\ket{11}_{out}
\end{aligned}
\end{equation}
Therefore, the 16 elements of the unitary operator $\bar{U}$ from the $W$-matrix on-chip are:
\begin{equation}
    \bar{U}_{ij}=\rm perm\left(\begin{bmatrix}
        w_{s_i(1),s_j(1)} &w_{s_i(1),s_j(2)} &w_{s_i(1),s_j(3)}\\
        w_{s_i(2),s_j(1)} &w_{s_i(2),s_j(2)} &w_{s_i(2),s_j(3)}\\
        w_{s_i(3),s_j(1)} &w_{s_i(3),s_j(2)} &w_{s_i(3),s_j(3)}\\
    \end{bmatrix}\right),
\label{eq: 10}
\end{equation}
where $i,j\in\{1,2,3,4\}$ and $s_1 = (1,3,5), s_2=(1,4,5), s_3=(2,3,5), s_4=(2,4,5)$ with $\hat{a}_5$ the ancilla mode. The 16 equations are specifically:
\begin{align*}
    \bar{U}_{11} = \rm perm(\begin{bmatrix}
        w_{11} &w_{13} &w_{15}\\
        w_{31} &w_{33} &w_{35}\\
        w_{51} &w_{53} &w_{55}\\
    \end{bmatrix}),
    \bar{U}_{12} = \rm perm(\begin{bmatrix}
        w_{11} &w_{14} &w_{15}\\
        w_{31} &w_{34} &w_{35}\\
        w_{51} &w_{54} &w_{55}\\
    \end{bmatrix}),
    \bar{U}_{13} = \rm perm(\begin{bmatrix}
        w_{12} &w_{13} &w_{15}\\
        w_{32} &w_{33} &w_{35}\\
        w_{52} &w_{53} &w_{55}\\
    \end{bmatrix}),
    \bar{U}_{14} = \rm perm(\begin{bmatrix}
        w_{12} &w_{14} &w_{15}\\
        w_{32} &w_{34} &w_{35}\\
        w_{52} &w_{54} &w_{55}\\
    \end{bmatrix})\\
    \bar{U}_{21}=\rm perm(\begin{bmatrix}
        w_{11} &w_{13} &w_{15}\\
        w_{41} &w_{43} &w_{45}\\
        w_{51} &w_{53} &w_{55}\\
    \end{bmatrix}),
    \bar{U}_{22} = \rm perm(\begin{bmatrix}
        w_{11} &w_{14} &w_{15}\\
        w_{41} &w_{44} &w_{45}\\
        w_{51} &w_{54} &w_{55}\\
    \end{bmatrix}),
    \bar{U}_{23} = \rm perm(\begin{bmatrix}
        w_{12} &w_{13} &w_{15}\\
        w_{42} &w_{43} &w_{45}\\
        w_{52} &w_{53} &w_{55}\\
    \end{bmatrix}),
    \bar{U}_{24} = \rm perm(\begin{bmatrix}
        w_{12} &w_{14} &w_{15}\\
        w_{42} &w_{44} &w_{45}\\
        w_{52} &w_{54} &w_{55}\\
    \end{bmatrix})\\
    \bar{U}_{31} = \rm perm(\begin{bmatrix}
        w_{21} &w_{23} &w_{25}\\
        w_{31} &w_{33} &w_{35}\\
        w_{51} &w_{53} &w_{55}\\
    \end{bmatrix}),
    \bar{U}_{32} = \rm perm(\begin{bmatrix}
        w_{21} &w_{24} &w_{25}\\
        w_{31} &w_{34} &w_{35}\\
        w_{51} &w_{54} &w_{55}\\
    \end{bmatrix}),
    \bar{U}_{33} = \rm perm(\begin{bmatrix}
        w_{22} &w_{23} &w_{25}\\
        w_{32} &w_{33} &w_{35}\\
        w_{52} &w_{53} &w_{55}\\
    \end{bmatrix}),
    \bar{U}_{34} = \rm perm(\begin{bmatrix}
        w_{22} &w_{24} &w_{25}\\
        w_{32} &w_{34} &w_{35}\\
        w_{52} &w_{54} &w_{55}\\
    \end{bmatrix})\\
    \bar{U}_{41} = \rm perm(\begin{bmatrix}
        w_{21} &w_{23} &w_{25}\\
        w_{41} &w_{43} &w_{45}\\
        w_{51} &w_{53} &w_{55}\\
    \end{bmatrix}),
    \bar{U}_{42} = \rm perm(\begin{bmatrix}
        w_{21} &w_{24} &w_{25}\\
        w_{41} &w_{44} &w_{45}\\
        w_{51} &w_{54} &w_{55}\\
    \end{bmatrix}),
    \bar{U}_{43} = \rm perm(\begin{bmatrix}
        w_{22} &w_{23} &w_{25}\\
        w_{42} &w_{43} &w_{45}\\
        w_{52} &w_{53} &w_{55}\\
    \end{bmatrix}),
    \bar{U}_{44} = \rm perm(\begin{bmatrix}
        w_{22} &w_{24} &w_{25}\\
        w_{42} &w_{44} &w_{45}\\
        w_{52} &w_{54} &w_{55}\\
    \end{bmatrix})\\
\end{align*}
The aim is to make $\bar{U}$ approach $U_{\rm CNOT}$. The CNOT unitary is 
\begin{equation}
    U_{\rm CNOT}=\begin{bmatrix}
        1 &0 &0 &0\\
        0 &1 &0 &0\\
        0 &0 &0 &1\\
        0 &0 &1 &0
    \end{bmatrix}
\end{equation}
The trained $W$ is 
\begin{equation}
    W = \begin{bmatrix}
        -0.2999 &0 &0.4677 &0.4677 &-1.1596\\
        0 &1.3682 &0 &0 &0\\
        0.8775 &0 &0.5341 &-0.8341 &-0.3480\\
        0.8775 &0 &-0.8341 &0.5341 &-0.3480\\
        -0.4921 &0 &-0.8199 &-0.8199 &-0.5341
    \end{bmatrix}
\end{equation}
The normalized matrix $\tilde{W} = W / \Vert W \Vert$ happens to be a unitary matrix with $\tilde{W} \tilde{W}^\dagger = \tilde{W}^\dag \tilde{W} = I$.

\section*{Appendix F: KL Divergence}
During and after the training, we calculate the Kullback Leibler (KL) divergence between theory/experimental transition probability distributions to measure the accuracy of the stochastic process modeling, which is given by
\begin{equation}
    D_{KL} = \sum_{x,i} P(S_i)P(x,S_j|S_i){\rm log}\left(\frac{P(x,S_j|S_i)}{\hat{P}(x,S_j|S_i)}\right).
\end{equation}

\section*{Appendix G: Dual Poisson process}
In the Dual Poisson process, the system first randomly turns on one of the two quantum channels with probability $p$ and $1-p$. Each channel can survive with probability $q_1$ and $q_2$.  
If the channel does not survive, the system resets and randomly turns on one of the quantum channels again. 
The survival event is recorded as 0 while the reset is recorded as 1. 
The process can be fully characterized by the survival probability $\Phi(k)$ that the process generates $k$ continuous 0s since the last output 1: $\Phi(k)=pq_1^k+\bar{p}q_2^k$, $\bar{p}=1-p$. The Dual Poisson process contains infinitely causal states $\{S_k|k\geq0\}$. 
The transition probability between casual states can be induced from the survival probability: 
\begin{equation}
\begin{aligned}
    &P(S_{k+1},0|S_k)=\frac{\Phi(k+1)}{\Phi(k)}, \\
    &P(S_0,1|S_k)=1-\frac{\Phi(k+1)}{\Phi(k)}.
\end{aligned}
\end{equation}
The quantum model works by encoding the causal state $S_k$ into the quantum state
\begin{equation}
\ket{\sigma_k}=\frac{\sqrt{pq_1^k}+ig\sqrt{\bar{p}q_2^k}}{\sqrt{\Phi(k)}}\ket{0}+\frac{i\sqrt{(1-g^2)\bar{p}q_2^k}}{\sqrt{\Phi(k)}}\ket{1},
\end{equation}
where $g=\frac{\sqrt{(1-q_1)(1-q_2)}}{1-\sqrt{q_1q_2}}$. Our target is to find the unitary that replicates the transition logic between casual states as
\begin{equation}
U\ket{\sigma_k}\ket{0}=\sqrt{\frac{\Phi(k+1)}{\Phi(k)}}\ket{\sigma_{k+1}}\ket{0}+\sqrt{1-\frac{\Phi(k+1)}{\Phi(k)}}\ket{\sigma_0}\ket{1}.
\end{equation}

\section*{Appendix H: Classical \& Quantum Models}

The classical models predict a stochastic process's future $\overrightarrow{x} := x_{0:\infty}$ based on the past information $\overleftarrow{x}:=x_{-\infty:0}$, $P(\overrightarrow{x}| \overleftarrow{x})$. 
Since storing the entire past is memory-consuming, the classical models encode the past into the past of a stochastic process into classical states $S_i$. 
The future is then generated according to a set of transition rules.
At each time step, the memory system at state $S_i$ transits to state $S_j$ and emits an output $x$ with probability $P(x,S_j|S_i)$.
Sequential transitions between different memory states generate a sequence of outputs.
Since memorizing the classical state is sufficient for generating future statistics, the amount of memory is naturally quantified by the number of classical states,
\begin{equation}
D_q = \#\{S_i \}.
\end{equation}
Meanwhile, as certain classical states are more likely to occur than others, an alternative way to quantify the amount of memory is the Shannon entropy of the memory states,
\begin{equation}
    C_c := H(\pi) = \sum_i -\pi_i \log_2 \pi_i. 
\end{equation}
where $\pi_i$ is the probability that each of the classical states $S_i$ occurs.

The quantum models encode the classical states $S_i$ into quantum states $\ket{\sigma_i}$.
At each timestep, the quantum models couple the memory system with an ancillary system in the vacuum state $\ket{0}$ by a unitary operator.
\begin{equation}
    U\ket{\sigma_i}\ket{0} = \sum_x \sqrt{P(x,S_j|S_i)}e^{i\phi} \ket{\sigma_j}\ket{0}
\end{equation}
The entropy of the quantum memory states is the von Neumann entropy of the mixed memory states,
\begin{equation}
    C_q := S(\rho) = \sum_i S(\pi_i \ketbra{\sigma_i}{\sigma_i}) 
\end{equation}
Since the quantum memory states are generally non-orthogonal, the entropy of quantum memory states $C_q$ is less than or equal to the entropy of the classical memory states,
\begin{equation}
    C_q \leq C_c
\end{equation}

\begin{figure}[h]
\centering
\includegraphics[width=0.75\textwidth]{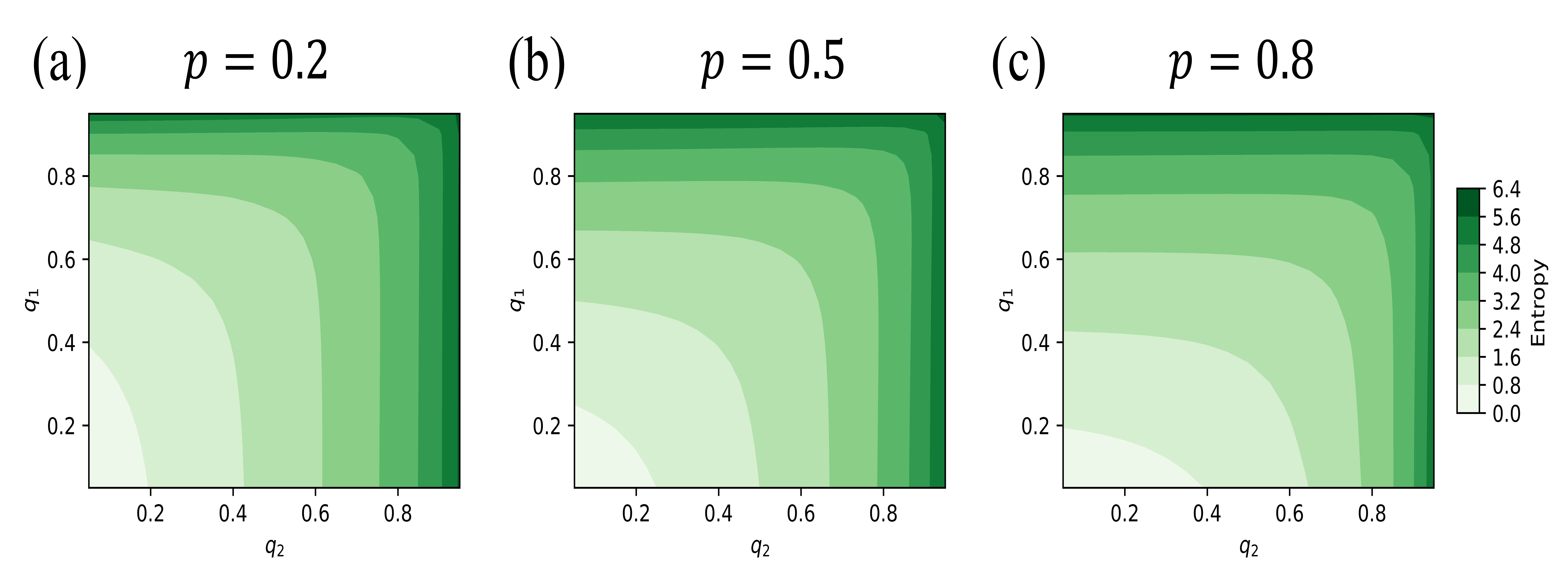}
\caption{The classical entropy of the dual poison process.}
\label{fig:classical}
\vspace{-0.3cm}
\end{figure}

The theoretical quantum entropy of the dual poisson process calculated is shown in the main text Figs.5a-c. The classical entropy entropy is shown in Fig.~\ref{fig:classical}. The results proves the reduced entropy $C_q<C_c$ for any parameters $p$, $q_1$ and $q_2$. 

\bibliography{n}